\documentclass[reprint,superscriptaddress]{revtex4-1}
\usepackage{graphicx}
\usepackage{amssymb}
\usepackage{amsmath}
\usepackage{epsfig}
\usepackage{chemarr}
\usepackage{url}
\usepackage{natbib}
\usepackage{dcolumn}
\usepackage{bm}
\usepackage{subfigure}
\usepackage{color}

\begin{document}

\title{Non-equilibrium statistical mechanics of continuous attractors}
\author{Weishun Zhong}
\affiliation{James Franck Institute, University of Chicago, Chicago, IL}
\affiliation{Department of Physics, MIT, Cambridge, MA}
\author{Zhiyue Lu}
\affiliation{James Franck Institute, University of Chicago, Chicago, IL}
\author{David J Schwab}
\email{dschwab@gc.cuny.edu}
\affiliation{Initiative for the Theoretical Sciences, CUNY Graduate Center}
\affiliation{Center for the Physics of Biological Function, Princeton \& CUNY}
\author{Arvind Murugan}
\email{amurugan@uchicago.edu}
\affiliation{Department of Physics and the James Franck Institute, University of Chicago, Chicago, IL}
\begin{abstract}
Continuous attractors have been used to understand recent neuroscience experiments where persistent activity patterns encode internal representations of external attributes like head direction or spatial location. However, the conditions under which the emergent bump of neural activity in such networks can be manipulated by space and time-dependent external sensory or motor signals are not understood. Here, we find fundamental limits on how rapidly internal representations encoded along continuous attractors can be updated by an external signal.  We apply these results to place cell networks to derive a velocity-dependent non-equilibrium memory capacity in neural networks.
\end{abstract}

\maketitle


Dynamical attractors have found much use in neuroscience as models for carrying out computation and signal processing \cite{Poucet2005-nm}. While point-like neural attractors and analogies to spin glasses have been widely explored \cite{Hopfield1982-fb,Amit1985-ls}, an important class of experiments are explained by `continuous attractors' where the collective dynamics of strongly interacting neurons stabilizes a low-dimensional family of activity patterns. Such continuous attractors have been invoked to explain experiments on motor control based on path integration \cite{Seung1996-ck,Seung2000-bk}, head direction \cite{Kim2017-zs} control, 
spatial representation in grid or place cells \cite{Yoon2013-nl,OKeefe1971-jt,Colgin2010-rd,Wills2005-hz,Wimmer2014-og, Pfeiffer2013-qn}, amongst other information processing tasks \cite{Hopfield2015-wt,Roudi2007-pm,Latham2003-fr,Burak2012-bu}. 

These continuous attractor models are at the fascinating intersection of dynamical systems and neural information processing. The neural activity in these models of strongly interacting neurons is described by an emergent collective coordinate \cite{Yoon2013-nl,Wu2008-iw,Amari1977-hf}. This  collective coordinate stores an internal representation \cite{Sontag2003-lk,Erdem2012-uf} of the organism's state in its external environment, such as position in space \cite{Pfeiffer2013-qn,McNaughton2006-xq} or head direction \cite{Seelig2015-uu}.

However, such internal representations are useful only if they can be driven and updated by external signals that provide crucial motor and sensory input  \cite{Hopfield2015-wt, Pfeiffer2013-qn,Erdem2012-uf,Hardcastle2015-as,Ocko2018-gv}.  Driving and updating the collective coordinate using external sensory signals opens up a variety of capabilities, such as path planning \cite{Ponulak2013-op,Pfeiffer2013-qn}, correcting errors in the internal representation or in sensory signals \cite{Erdem2012-uf,Ocko2018-gv}, and the ability to resolve ambiguities in the external sensory and motor input \cite{Hardcastle2015-as,Evans2016-pr,Fyhn2007-ys}.

In all of these examples, the functional use of attractors requires interaction between external signals and the internal recurrent network dynamics. However, with a few significant exceptions \cite{Wu2008-iw,Wu2005-sw,Monasson2014-nu,Burak2012-bu}, most theoretical work has either been in the limit of no external forces and strong internal recurrent dynamics, or in the limit of strong external forces where the internal recurrent dynamics can be ignored \cite{Moser2017-dj,Tsodyks1999-px}. 

Here, we study continuous attractors in neural networks subject to external driving forces that are neither small relative to internal dynamics, nor adiabatic. We show that the physics of the emergent collective coordinate sets limits on the maximum speed with which the internal representation can be updated by external signals.

Our approach begins by deriving simple classical and statistical laws satisfied by the collective coordinate of many neurons with strong, structured interactions that are subject to time-varying external signals, Langevin noise, and quenched disorder. Exploiting these equations, we demonstrate two simple principles; (a) an `equivalence principle' that 
predicts how much the internal representation lags a rapidly moving external signal, (b)  under externally driven conditions, quenched disorder in network connectivity can be modeled as a state-dependent effective temperature. 
Finally, we apply these results to place cell networks and derive a non-equilibrium driving-dependent memory capacity, complementing numerous earlier works on memory capacity in the absence of external driving.

\subsection*{Collective coordinates in continuous attractors}
We study $N$ interacting neurons following the formalism presented in \cite{Hopfield2015-wt},
\begin{equation}
\label{eqn:eom_zero}
\frac{di_n}{dt} = -\frac{i_n}{\tau} + \sum_{k = 1}^{N} J_{nk}f(i_k) + I^{ext}_n(t) +\eta_{int}(t),
\end{equation}
where $f(i_k) = (1+e^{-i_k/i_0})^{-1}$ is the neural activation function that represents the firing rate of neuron $k$, and $i_n$ is an internal excitation level of neuron $n$ akin to the membrane potential. We consider synaptic connectivity matrices with two distinct components, 
\begin{equation}
J_{ij} = J_{ij}^{0}+ J_{ij}^{d}.
\label{eqn:JijBreakdown}
\end{equation}

\begin{figure}
\includegraphics[width=\linewidth]{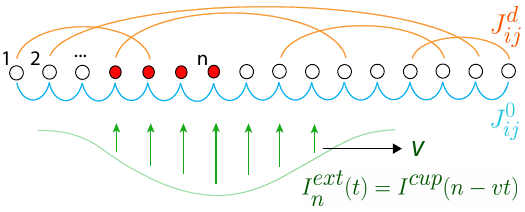}
\caption{ 
The effective dynamics of neural networks implicated in head direction and spatial memory is described by a continuous attractor. Consider $N$ neurons connected in a 1-D topology, with local excitatory connections between $p$ nearest neighbors (blue), global inhibitory connections (not shown), and random long-range disorder (orange). Any activity pattern quickly condenses into a `droplet' of contiguous firing neurons (red) of characteristic size; the droplet center of mass $\bar{x}$ is a collective coordinate parameterizing a continuous attractor. The droplet can be driven by space and time-varying external currents $I^{ext}_{n}(t)$ (green). \label{fig:schematics}} 
\end{figure}

As shown in Fig.\ref{fig:schematics}, $J_{ij}^{0}$ encodes the continuous attractor. We will focus on $1$-D networks with $p$-nearest neighbor excitatory interactions to keep bookkeeping to a minimum: $J^{0}_{ij} = J(1 - \epsilon)$ if neurons $|i-j| \leq p$, and $J_{ij}^{0} = - J \epsilon$ otherwise. The latter term, $- J \epsilon$, with $0\leq \epsilon \leq 1$, represents long-range, non-specific inhibitory connections 
as frequently assumed in models of place cells \cite{Monasson2013-kx,Hopfield2010-lf}, head direction cells \cite{Chaudhuri2016-lh} and other continuous attractors \cite{Seung2000-bk,Burak2012-bu}. 

The disorder matrix $J_{ij}^{d}$ represents random long-range connections, a form of quenched disorder \cite{Sebastian_Seung1998-vm,Kilpatrick2013-go}. 
Finally, $I_n^{ext}(t)$ represents external driving currents from e.g. sensory and motor input possibly routed through other regions of the brain. The Langevin noise $\eta_{int}(t)$ represents private noise internal to each neuron \cite{Lim2012-ek,Burak2012-bu} with $\langle \eta_{int}(t) \eta_{int}(0) \rangle = C_{int}\delta(t)$.

A neural network with $p$-nearest neighbor interactions like Eqn.\eqref{eqn:eom_zero} qualitatively resembles a similarly connected network of Ising spins; the inhibitory connections impose a (soft) constraint on the number of neurons that can be firing at any given time and hence \cite{Monasson2013-pn} similar to working at fixed magnetization in an Ising model. 
At low noise, the activity in such a system will condense \cite{Monasson2013-kx,Hopfield2010-lf} to a localized `droplet', since interfaces between firing and non-firing neurons are penalized by $J(1-\epsilon)$. 
The center of mass of such a droplet, $\bar{x} \equiv \frac{\sum_n n f(i_n) }{\sum_n f(i_n)}$ is an emergent collective coordinate that approximately describes the stable low-dimensional neural activity patterns of these $N$ neurons. Fluctuations about this coordinate have been extensively studied \cite{Wu2008-iw,Burak2012-bu,Hopfield2015-wt,Monasson2014-nu}.

\subsection*{Space and time dependent external signals}

We focus on how space and time-varying external signals, modeled here as external currents $I^{ext}_n(t)$ can drive and reposition the droplet along the attractor. We will be primarily interested in a cup-shaped current profile that moves at a constant velocity $v$, i.e., $I^{ext}_n(t) = I^{cup}(n - vt)$ where 
$I^{cup}(n) = d(w-|n|), n \in [-w,w]$, $I^{cup}(n) = 0$ otherwise. Such a localized time-dependent drive could represent landmark-related sensory signals \cite{Hardcastle2015-as} when a rodent is traversing a spatial environment at velocity $v$, or signals that update the internal representation of head direction \cite{Seelig2015-uu}.

In addition to such positional information, continuous attractors often also receive velocity information \cite{Major2004-ku,McNaughton2006-xq,Seelig2015-uu,Ocko2018-gv}; such signals are modeled \cite{Burak2009-ch,Hopfield2010-lf} as a time-independent anti-symmetric $A^{0}_{ij}$ added on to $J^{0}_{ij} \to J^0_{ij} + A^{0}_{ij}$ that `tilts' the continuous attractor, so the droplet moves with a velocity proportional to $A^0_{ij}$. 

Such velocity integration (or `dead-reckoning') will inevitably accumulate errors that are then corrected using direct positional information modeled by $I^{ext}_n(t)$ \cite{Hardcastle2015-as}. In the Appendix, we find that in the presence of $A_{ij}$, the velocity $v$ of $I^{ext}(t)$ can be interpreted as the difference in velocity implied by positional and velocity information, which has been manipulated in virtual reality experiments \cite{Major2004-dz,Colgin2010-rd,Seelig2015-uu, Ocko2018-gv, Campbell2018-wt}. Therefore, for simplicity here we set $A_{ij}=0$.

The effective dynamics of the collective coordinate $\bar{x}$ in the presence of currents $I^{ext}_n(t)$ can be obtained by computing the effective force on the droplet of finite size. We find that (see Appendix)
\begin{equation}
\gamma \dot{\bar{x}} = - \partial_{\bar{x}} V^{ext}(\bar{x},t), 
\end{equation}
where $V^{ext}(\bar{x},t)$ is a piecewise quadratic potential $V^{cup}(\bar{x}-vt)$ for currents $I^{ext}_n(t) = I^{cup}(n - vt)$, and $\gamma$ is the effective drag coefficient of the droplet. (Here, we neglect rapid transients of timescale $\tau$ \cite{Wu2008-iw}.) 

The strength of the external signal is set by the depth $d$ of the cup $I^{cup}(n)$. Previous studies have explored the $d = 0$ case, i.e., undriven diffusive dynamics of the droplet \cite{Burak2012-bu,Monasson2014-nu, Monasson2013-pn,Monasson2015-nl}. Studies have also explored large $d$ \cite{Hopfield2015-wt} when the internal dynamics can be ignored. In fact, as shown in the Appendix, we find a threshold signal strength $d_{max}$ beyond which the external signal destabilizes the droplet,  instantly `teleporting' the droplet from any distant location to the cup without continuity along the attractor, erasing any prior positional information held in the internal representation.

\begin{figure}
\includegraphics[width=1\linewidth]{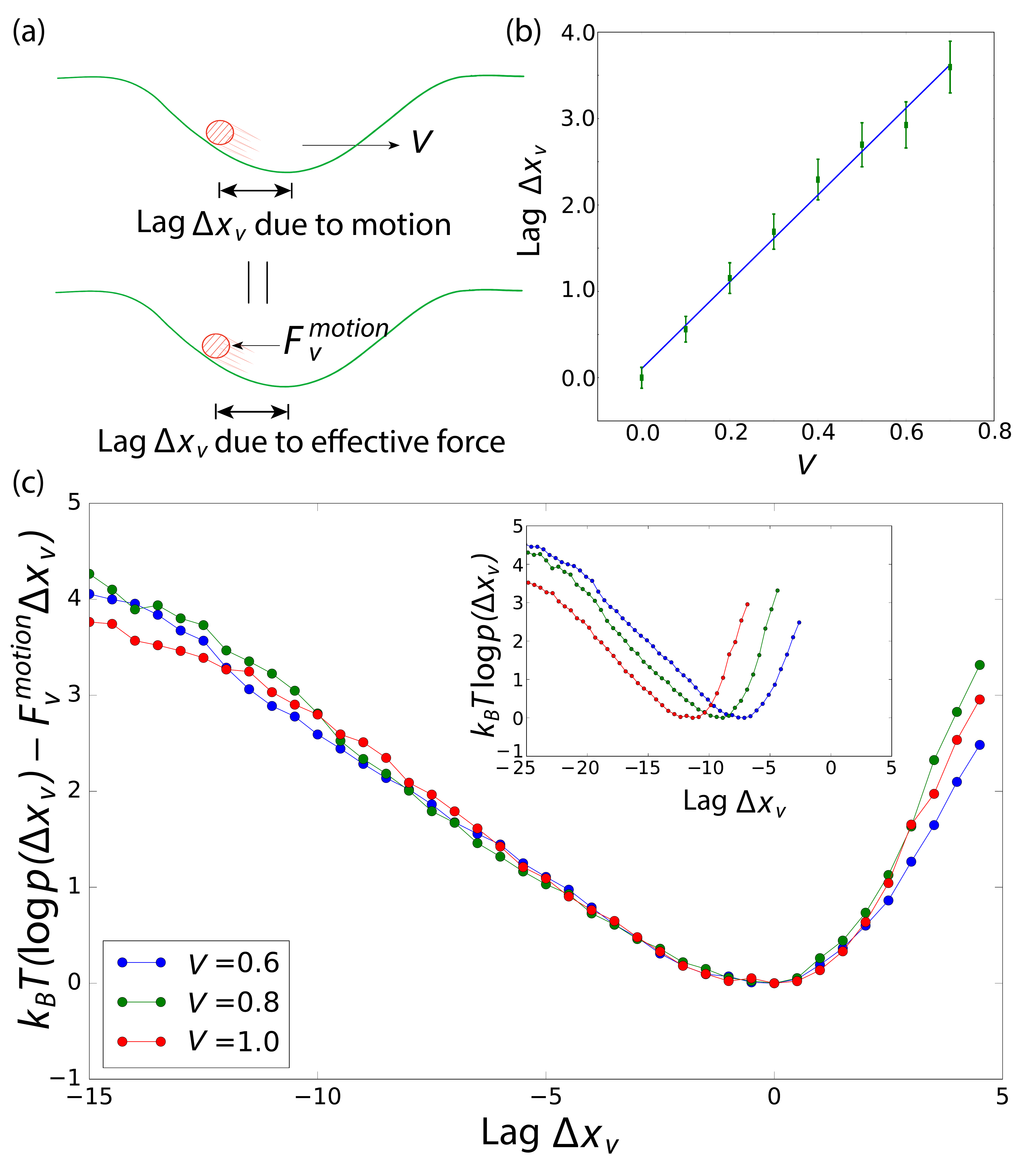}
\caption{(a) The mean position and fluctuations of the droplet driven by currents $I_{n}^{ext} = I^{cup}(n-vt)$ are described by an `equivalence' principle; in a frame co-moving with $I_n^{cup}(t)$ with velocity $v$, we simply add an effective force $F^{motion}_{v} = \gamma v$ where $\gamma$ is a drag coefficient. (b) This prescription correctly predicts that the droplet lags the external driving force by an amount linearly proportional to velocity $v$, as seen in simulations. (c) Fluctuations of the driven droplet's position, due to internal noise in neurons, are also captured by the equivalence principle. If $p(\Delta x_v)$ is the probability of finding the droplet at a lag $\Delta x_v$, we find that $k_B T \log p(\Delta x_v) - k_B T F^{motion}_{v} \Delta x_v$ is independent of velocity and can be collapsed onto each other (with fitting parameter $T$).  (Inset: $\log p(\Delta x_v)$ before subtracting $F^{motion}_{v} x$.)
\label{fig:equiv principle}}
\end{figure}

We focus here on $d < d_{max}$, a regime with continuity of internal representations. Such continuity is critical for many applications such as path planning \cite{Ponulak2013-op,Pfeiffer2013-qn,Erdem2012-uf} and resolving local ambiguities position within the global context \cite{Hardcastle2015-as,Evans2016-pr,Fyhn2007-ys}. In this regime, the external signal updates the internal representation with finite `gain' \cite{Fyhn2007-ys} and can thus fruitfully combine information in both the internal representation and the external signal. Other applications that simply require short-term memory storage of a strongly fluctuating variable may not require this continuity restriction.

\subsubsection*{Equivalence principle}
We first consider driving the droplet in a network at constant velocity $v$ using an external current $I^{ext}_n = I^{cup}(n - vt)$. We allow for Langevin noise but no disorder in the couplings $J^{d} = 0$ in this section. For very slow driving ($v \to 0$), the droplet will settle into and track the bottom of the cup. When driven at a finite velocity $v$, the droplet cannot stay at the bottom since there is no net force exerted by the currents $I^{ext}_n$ at that point.  

Instead, the droplet must lag the bottom of the moving external drive by an amount $ \Delta x_v  = \bar{x} - v t$ such that the slope of the potential $V^{cup}$ provides an effective force $F_v^{motion} \equiv \gamma v$ needed to keep the droplet in motion at velocity $v$. That is, the lag $\Delta x_v$ when averaged over a long trajectory, must be,
\begin{equation} 
-\partial_{\bar{x}} V^{cup}(\langle \Delta x_v \rangle)  = F_v^{motion} \equiv \gamma v.
\label{eqn:equivalance}
\end{equation}
This equation is effectively an `equivalence' principle for over-damped motion -- in analogy with inertial particles accelerated in a potential, the droplet lags to a point where the slope of the driving potential provides sufficient force to keep the droplet in motion at that velocity. Fig.~\ref{fig:equiv principle}b verifies that the average lag $\langle \Delta x_v \rangle$ depends on velocity in a way described by Eqn.~\ref{eqn:equivalance}.

In fact, the above `equivalence' principle goes beyond predicting the mean lag $\langle \Delta x_v \rangle$; the principle also correctly predicts the entire distribution $p(\Delta x_v)$ of fluctuations of the lag $\Delta x_v$ due to Langevin noise; see Fig.\ref{fig:equiv principle}c. By binning the lag $\Delta x_v(t)$ for trajectories of the droplet obtained from repeated numerical simulations, we determined $p(\Delta x_v)$, the occupancy of the droplet in the moving frame of the drive. We find that $\log p(\Delta x_v)$ for different velocities corresponds to the same quadratic potential $V^{cup}$ plus a velocity-dependent linear potential, $-F^{motion}_v \Delta x_v$, in agreement with the equivalence principle. That is,
\begin{equation}
\label{eqn:fluc lag}
k_B T \log p(\Delta x_v) = -(V^{{cup}}(\Delta x_v) - F^{{motion}}_v \Delta x_v),
\end{equation}
for some effective temperature scale $T$ for the collective coordinate $\bar{x}$, ultimately set by $\eta_{int}(t)$. (See Appendix.) As a result, the $\log p(\Delta x_v)$ for different velocities collapse onto each other upon subtracting the linear potential due to the motion force, as shown in Fig.\ref{fig:equiv principle}c. 

In summary, in the co-moving frame of the driving signal, the droplet's position $\Delta x_v$ fluctuates as if it were in thermal equilibrium in the modified potential $V^{eff} = V^{cup} - F^{motion}_v \Delta x_v$.  

\subsection*{Speed limits on updates of internal representation}
These results for the distribution of the lag $\Delta x_v$, captured by a simple `equivalence principle', imply a striking restriction on the speed at which external positional information can update the internal representation. A driving signal of strength $d$ cannot drive the droplet at velocities greater than some $v_{crit}$ if the predicted lag for $v > v_{crit}$ is larger than the cup. In the Appendix, we find $v_{crit} = 2d(w+R)/3\gamma$, where $2R$ is the droplet size.

Larger driving strength $d$ increases $v_{crit}$, but as was previously discussed, we require $d < d_{max}$ in order to retain continuity and stability of the internal representation, i.e. to prevent teleportation of the activity bump. Hence, we find an absolute upper bound on the fastest external signal that can be tracked by the internal dynamics of the attractor,

\begin{equation}
v^* = \kappa p J\gamma^{-1},
\label{eqn:vfund}
\end{equation}
where $p$ is the range of interactions, $J$ is the synaptic strength, $\gamma^{-1}$ is the mobility or inverse drag coefficient of the droplet, and $\kappa$ is a dimensionless $\mathcal{O}(1)$ number.


\begin{figure}
 \includegraphics[width=1\linewidth]{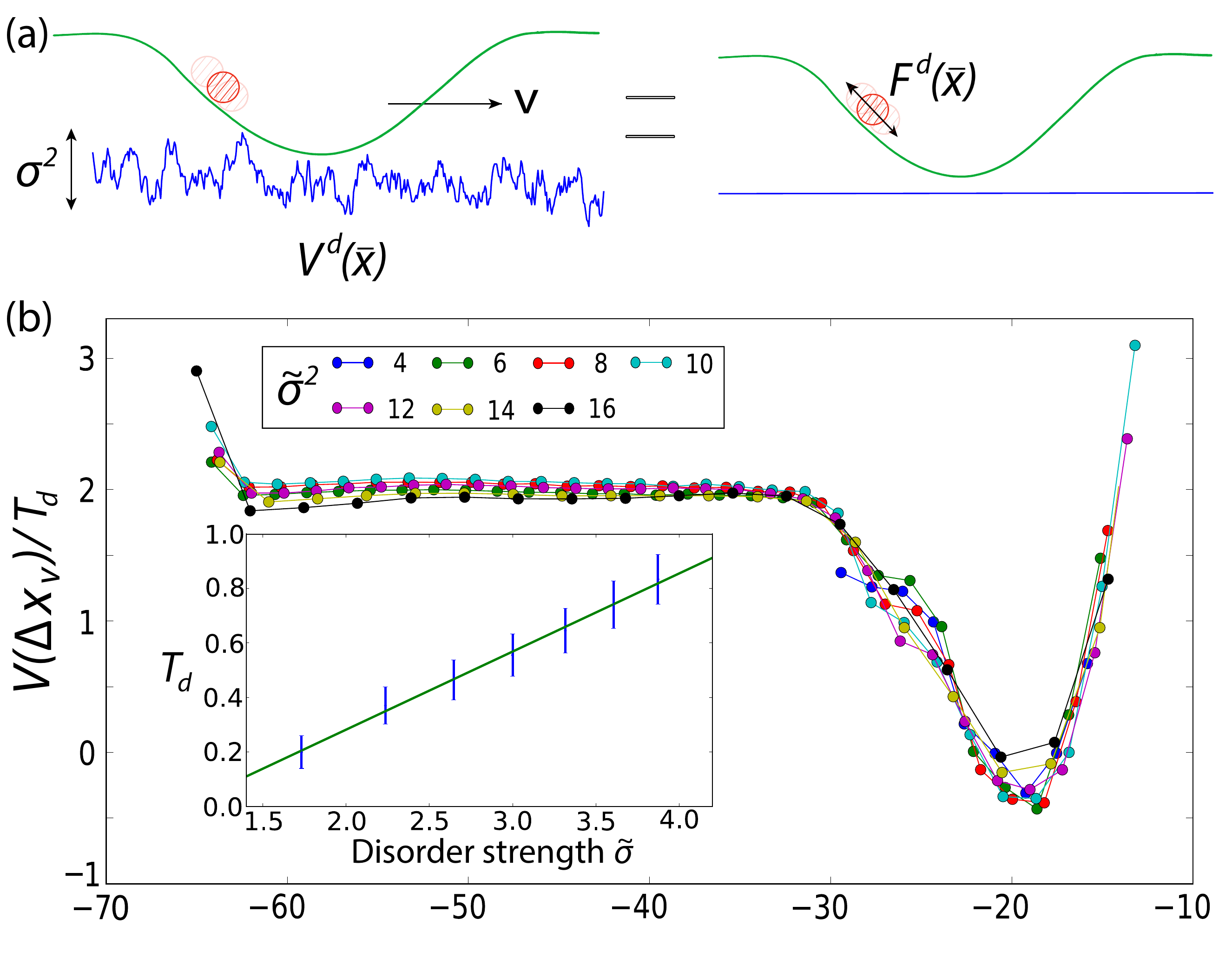}
\caption{Disorder in neural connectivity is well-approximated by an effective temperature $T_d$ for a moving droplet. (a) Long-range disorder breaks  the degeneracy of the continuous attractor, creating a rough landscape. A droplet moving at velocity $v$ in this rough landscape experiences random forces. (b) The fluctuations of a moving droplet's position, relative to the cup's bottom, can be described by an effective temperature $T_{d}$. We define a potential $V(\Delta x_v) = - k_B T_{d} \log p(\Delta x_v)$ where $p(\Delta x_v)$ is the probability of the droplet's position fluctuating to a distance $\Delta x_v$ from the peak external current. We find that $V(\Delta x_v)$ corresponding to different amounts of disorder $\tilde{\sigma}^2$ (where $\tilde{\sigma}^2$ is the average number of long-ranged disordered connections per neuron in units of $2p$), can be collapsed by the one fitting parameter $T_{d}$. (inset) $T_{d}$ is linearly proportional to the strength of disorder $\tilde{\sigma}$. \label{fig:Tbumpy}}
\end{figure}

\begin{figure}
\includegraphics[width=1\linewidth]{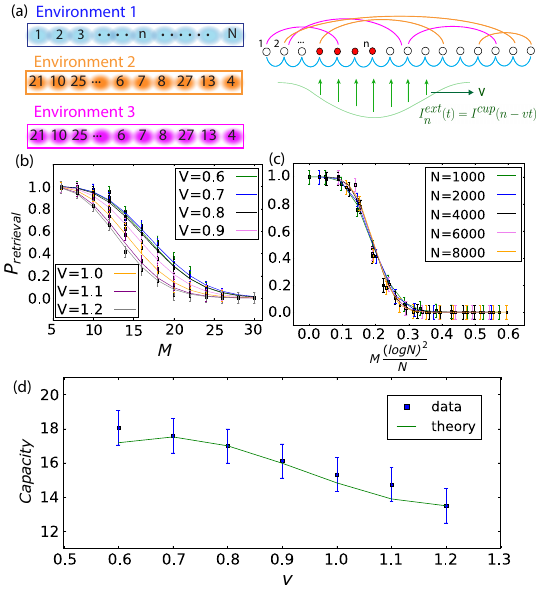}
\caption{Non-equilibrium capacity of place cell networks limits retrieval of spatial memories at finite velocity. (a) Place cell networks model the storage of multiple spatial memories in parts of the hippocampus by coding multiple continuous attractors in the same set of neurons. Neural connections encoding spatial memory 2,3,\ldots act like long range disorder for spatial memory 1. Such disorder, through an increased effective temperature, reduces the probability of tracking a finite velocity driving signal. (b)  The probability of successful retrieval, $P_{retrieval}$, decreases with the number of simultaneous memories $M$ and velocity $v$ (with $N=4000,p=10,\epsilon=0.35,\tau=1,J=100,d=10,w=30$ held fixed). (c) $P_{retrieval}$ simulation data collapses when plotted against $M/(N/(\log N)^2)$ (parameters same as (b) with $v=0.8$ held fixed and $N$ varies).
(d) The non-equilibrium capacity $M_c$ as a function of retrieval velocity $v$.\label{fig:capacity}}
\end{figure}

\subsection*{Disordered connections and effective temperature}

We now consider the effect of long-range quenched disorder $J_{ij}^{d}$ in the synaptic matrix \cite{Sebastian_Seung1998-vm,Kilpatrick2013-go}, which breaks the exact degeneracy of the continuous attractor, creating an effectively rugged landscape, $V^{d}(\bar x)$, 
as shown schematically in Fig. \ref{fig:Tbumpy} and computed in the Appendix. 
When driven by a time-varying external signal, $I^{ext}_i(t)$, the droplet now experiences a net potential $V^{ext}(\bar{x},t) + V^{d}(\bar{x})$. The first term causes motion with velocity $v$ and a lag predicted by the equivalence principle. 
The second term $V^{d}(\bar{x})$ is difficult to handle in general. However, for sufficiently large velocities $v$, we find that the effect of $V^{d}(\bar{x})$ can be modeled as effective Langevin white noise. To see this, note that $V^{d}(\bar{x})$ is uncorrelated on length scales larger than the droplet size; hence for large enough droplet velocity $v$, the forces $F^{d}(t) \equiv -\partial_{\bar{x}}V^d\vert_{\bar{x} = \bar{x}(t)}$ due to disorder are effectively random and uncorrelated in time. More precisely, let $\sigma^2 = \text{Var}(V^d(\bar{x}))$. In the Appendix, we compute $F^{d}(t)$ and show that $F^d(t)$ has an auto-correlation time, 
$\tau_{cor} = 2R/v$ due to the finite size of the droplet.  

Thus, on longer timescales, $F^{d}(t)$ is uncorrelated and can be viewed as Langevin noise for the droplet center of mass $\bar{x}$, associated with a disordered-induced temperature $T_d$. Through repeated simulations with different amounts of disorder $\sigma^2$, we inferred the distribution $p(\Delta x_v)$ of the droplet position in the presence of such disorder-induced fluctuations; see Fig. \ref{fig:Tbumpy}. The data collapse in Fig. \ref{fig:Tbumpy}b confirms that the effect of disorder (of size $\sigma^2$) on a rapidly moving droplet can indeed by modeled by an effective disorder-induced temperature $T_{d} \sim \sigma \tau_{cor}$.
(For simplicity, we assume that internal noise $\eta_{int}$ in Eqn.\eqref{eqn:eom_zero} is absent here.)  

Thus, the disorder $J_{ij}^d$ effectively creates thermal fluctuations about the lag predicted by the equivalence principle; such fluctuations may carry the droplet out of the driving cup $I^{cup}(n-vt)$ and prevent successful update of the internal representation. We found that this effect can be quantified by a simple Arrhenius-like law, \begin{equation}
\label{eqn:DeltaE}
r \sim \exp(- \Delta E(v,d) / k_{B} T_{d})
\end{equation}

where $\Delta E(v,d)$ is the energy gap between where the droplet sits in the drive and the escape point, predicted by the equivalence principle, and $T_d$ is the disorder-induced temperature.  Thus, given a network of $N$ neurons, the probability of an external drive moving the droplet successfully across the network is proportional to $\exp(- r N)$.

\subsection*{Memory capacity of driven place cell networks}

The capacity of a neural network to encode multiple memories has been studied in numerous contexts since Hopfield's original work \cite{Hopfield1982-fb}. While specifics differ \cite{Amit1985-as,Battaglia1998-eg,Monasson2013-pn,Hopfield2010-lf}, the capacity is generally set by the failure to retrieve a specific memory because of the effective disorder in neural connectivity due other stored memories.

However, these works on capacity do not account for non-adiabatic external driving. Here, we use our results to determine the capacity of a place cell network \cite{OKeefe1971-jt,Battaglia1998-eg,Monasson2013-pn} to both encode and manipulate memories of multiple spatial environments at a finite velocity. Place cell networks \cite{ Tsodyks1999-px,Monasson2013-kx, Monasson2015-nl, Monasson2014-nu, Monasson2013-pn} encode memories of multiple spatial environments as multiple continuous attractors in one network. Such networks have been used to describe recent experiments on place cells and grid cells in the hippocampus \cite{Yoon2013-nl, Hardcastle2015-as,Moser2014-lw}.

In experiments that expose a rodent to different spatial environments $\mu = 1,\ldots M$ \cite{Alme2014-qc, Moser2017-dj, Kubie1991-ge}, the same place cells $i=1,\ldots N$ are seen having `place fields' in different spatial arrangements $\pi^\mu(i)$ as seen in Fig.\ref{fig:capacity}A, where $\pi^\mu$ is a permutation specific to environment $\mu$. Consequently, Hebbian plasticity suggests that each environment $\mu$ would induce a set of synaptic connections $J_{ij}^{\mu}$ that corresponds to the place field arrangement in that environment; i.e., $J_{ij}^{\mu} = J(1-\epsilon)$ if $|\pi^{\mu}(i) - \pi^{\mu}(j)| < p $. That is, each environment corresponds to a $1$-D network when the neurons are laid out in a specific permutation $\pi^\mu$. The actual network has the sum of all these connections $J_{ij} = \sum_{\mu=1}^M J_{ij}^\mu$ over the $M$ environments the rodent is exposed to.

While $J_{ij}$ above is obtained by summing over $M$ structured environments, from the perspective of, say, $J_{ij}^{1}$, the remaining $J_{ij}^{\mu}$ look like long-range disordered connections. We will assume that the permutations $\pi^{\mu}(i)$ corresponding to different environments are random and uncorrelated, a common modeling choice with experimental support \cite{Hopfield2010-lf,Monasson2015-nl, Monasson2014-nu,Alme2014-qc,Moser2017-dj}. Without loss of generality, we assume that $\pi^{1}(i) = i$ (blue environment in Fig.\ref{fig:capacity}.) Thus, $J_{ij} = J_{ij}^{1} + J_{ij}^{d}, J_{ij}^{d} = \sum_{\mu = 2}^N J_{ij}^{\mu}$. The disordered matrix $J^{d}_{ij}$ then has an effective variance $\sigma^2 \sim (M-1)/N$. Hence, we can apply our previous results to this system. 
Now consider driving the droplet with velocity $v$ in Environment 1 using external currents. The probability of successfully updating the internal representation over a distance $L$ is given by $P_{retrieval} = e^{- rL/v}$, where $r$ is given by Eqn.\eqref{eqn:DeltaE}.

In the thermodynamic limit $N \to \infty$, with $w, p, L/N$ held fixed, $P_{retrieval}$ becomes a Heaviside step function $\Theta (M_c-M)$ at some critical value $M_c$ given by
\begin{equation}
\label{eqn:capacity}
M_c \sim \bigg[v\Delta E(v,d)\bigg]^2 \frac{N}{(\log N)^2}
\end{equation}
for the largest number of memories that can be stored and retrieved at velocity $v$. $\Delta E(v,d) = (4dw-3\gamma v-2dR)(-v\gamma+2dR)/4d $. Fig.\ref{fig:capacity} shows that our numerics agree well with this formula, showing a novel dependence of the capacity of a neural network on the speed of retrieval and the strength of the external drive.

In this paper, we found that the non-equilibrium statistical mechanics of a strongly interacting neural network can be captured by a simple equivalence principle and a disorder-induced temperature for the network's collective coordinate. Consequently, we were able to derive a velocity-dependent bound on the number of simultaneous memories that can be stored and retrieved from a network. 
Our approach used specific functional forms for, e.g., the current profile $I^{cup}(n-vt)$. However, our bound simply reflects the finite response time in moving emergent objects, much like moving a magnetic domain in a ferromagnet using space and time varying fields. Thus we expect our bound to hold qualitatively for other related models \cite{Hopfield2015-wt}. Such general theoretical principles on driven neural networks are needed to connect to recent time-resolved experiments in neuroscience\cite{ Kim2017-zs,Turner-Evans2017-ch,Hardcastle2015-as} on the response of neural networks to dynamic perturbations.

\acknowledgements{We thank Jeremy England, Ila Fiete, John Hopfield, and Dmitry Krotov for discussions. AM and DS are grateful for support from the Simons Foundation MMLS investigator program. We acknowledge the University of Chicago Research Computing Center for support of this work.}

\pagebreak
\bibliographystyle{unsrt}
\bibliography{Paperpile}

\clearpage

\appendix

\section{Equations for the collective coordinate}
\label{app:eqn_cc}
As in the main text, we model $N$ interacting neurons as,

\begin{equation}
\begin{split}
\label{eqn:eom}
\frac{di_n}{dt} &= -\frac{i_n}{\tau} + \sum_{k = 1}^{N} J_{nk}f(i_k) + I_n^{ext}(t) + \eta^{int}_n(t),
\\
&\text{where} \; f(i) = \frac{1}{1+e^{-i/i_0}}.
\end{split}
\end{equation}
The synaptic connection between two different neurons $i,j$ is $J_{ij} = J(1 - \epsilon)$ if neurons $i$ and $j$ are separated by a distance of at most $p$ neurons, and $J_{ij}= - J \epsilon$ otherwise, and note that we set the self-interaction to zero. The internal noise is a white noise, $\langle \eta^{int}_n(t) \eta^{int}_n(0) \rangle = C_{int}\delta(t)$ with an amplitude $C_{int}$. $I_n^{ext}(t)$ are external  driving currents discussed below.

Such a quasi 1-d network with $p$-nearest neighbor interactions resembles a similarly connected network of Ising spins at fixed magnetization in its behavior; the strength of inhibitory connections $\epsilon$ constrains the total number of neurons $2R$ firing at any given time to $2R \sim p \epsilon^{-1}$. It was shown \cite{Hopfield2010-lf,Monasson2013-kx, Monasson2014-nu, Monasson2013-pn} that below a critical temperature $T$, the $w$ firing neurons condense into a contiguous droplet of neural activity, minimizing the total interface between firing and non-firing neurons. Such a droplet was shown to behave like an emergent quasi-particle that can diffuse or be driven around the continuous attractor. We define the center of mass of the droplet as,
\begin{equation}
\bar{x} \equiv \sum_n n f(i_n).
\end{equation}
The description of neural activity in terms of such a collective coordinate $\bar{x}$ greatly simplifies the problem, reducing the configuration space from the $2^N$ states for the $N$ neurons to $N$-state consists of the center of mass of the droplet along the continuous attractor \cite{Wu2008-iw}. Computational abilities of these place cell networks, such as spatial memory storage, path planning and pattern recognition, are limited to parameter regimes in which such a collective coordinate approximation holds (e.g., noise levels less than a critical value $T < T_c$) .

The droplet can be driven by external signals such as sensory or motor input or input from other parts of the brain. We model such external input by the currents $I_{n}^{ext}$ in Eqn.\ref{eqn:eom}; for example, sensory landmark-based input \cite{Hardcastle2015-as} when an animal is physically in a region covered by place fields of neurons $i, i+1,\ldots, i+z$, currents $I^{ext}_{i}$ through $I^{ext}_{i+z}$ can be expected to be high compared to all other currents $I^{ext}_j$. Other models of driving in the literature include adding an anti-symmetric component $A_{ij}$ to synaptic connectivities  $J_{ij}$ \cite{Ponulak2013-op}; we consider such a model in Appendix \ref{Aij}.

Let $\{i_k^{\bar{x}}\}$ denote the current configuration such that the droplet is centered at location $\bar{x}$. The Lyapunov function of the neural network is given by\cite{Hopfield2015-wt}, 

\begin{equation}
	\label{eqn:lyapunov}
	\begin{split}
		\mathcal{L}[\bar{x}] &\equiv \mathcal{L}[f(i_k^{\bar{x}})] \\
		&= \frac{1}{\tau}\sum_{k} \int_0^{f(i_k^{\bar{x}})}f^{-1}(x)dx\\
		& -\frac{1}{2} \sum_{n,k}J_{nk}f(i_k^{\bar{x}})f(i_n^{\bar{x}})- \sum_k f(i_k^{\bar{x}})I^{ext}_k(t).
	\end{split}
\end{equation}

In a minor abuse of terminology, we will refer to terms in the Lyapunov function as energies, even though energy is not conserved in this system. For future reference, we denote the second term $V_J(\bar{x}) = -1/2\sum_{nk}J_{nk}f(i_k^{\bar{x}})f(i_n^{\bar{x}})$, which captures the effect of network synaptic connectivities. Under the `rigid bump approximation' used in \cite{Hopfield2015-wt},i.e., ignoring fluctuations fo the droplet, we find,

\begin{eqnarray}
V_J(\bar{x}) &= -\frac{1}{2}\sum_{n,k} f(i_n^{\bar{x}}) J_{nk} f(i_k^{\bar{x}}) \\ &\approx -\frac{1}{2}\sum_{\substack{|n-\bar{x}| \le R, \\ |k-\bar{x}| \le R}} f(i_n^{\bar{x}}) J_{nk} f(i_k^{\bar{x}}).
\end{eqnarray}

For a quasi 1-d network with $p$-nearest neighbor interactions and no disorder, $V_J(\bar{x})$ is constant, giving a smooth continuous attractor. However, as discussed later, at the presence of disorder, $V_J(\bar{x})$ has bumps (i.e. quenched disorder) and is no longer a smooth continuous attractor.

To quantify the effect of the external driving, we write the third term in Eqn.\eqref{eqn:lyapunov},
 \begin{eqnarray}
 V^{ext}(\bar{x},t) &=& -\sum_k I^{ext}_k(t) f(i_k^{\bar{x}}) \\ 
 &\approx & -\sum_{|k-\bar{x}| < R} I^{ext}_k(t) f(i_k^{\bar{x}})
 \end{eqnarray}
 
Thus, the external driving current $I_n^{ext}(t)$ acts on the droplet through the Lyapunov function $V^{ext}(\bar{x},t)$. Hence we define
\begin{equation}
 F^{ext}(\bar{x},t) = -\partial_{\bar{x}} V^{ext}(\bar{x},t)  \\
\end{equation}
to be the external force acting on the droplet center of mass. 

\subsection*{Fluctuation and dissipation}
\label{fluctuation-dissipation}

We next numerically verify that the droplet obeys a fluctuation-dissipation-like relation by driving the droplet using external currents $I^{ext}$ and comparing the response to diffusion of the droplet in the absence of external currents.

\begin{figure}
\includegraphics[width=1\linewidth]{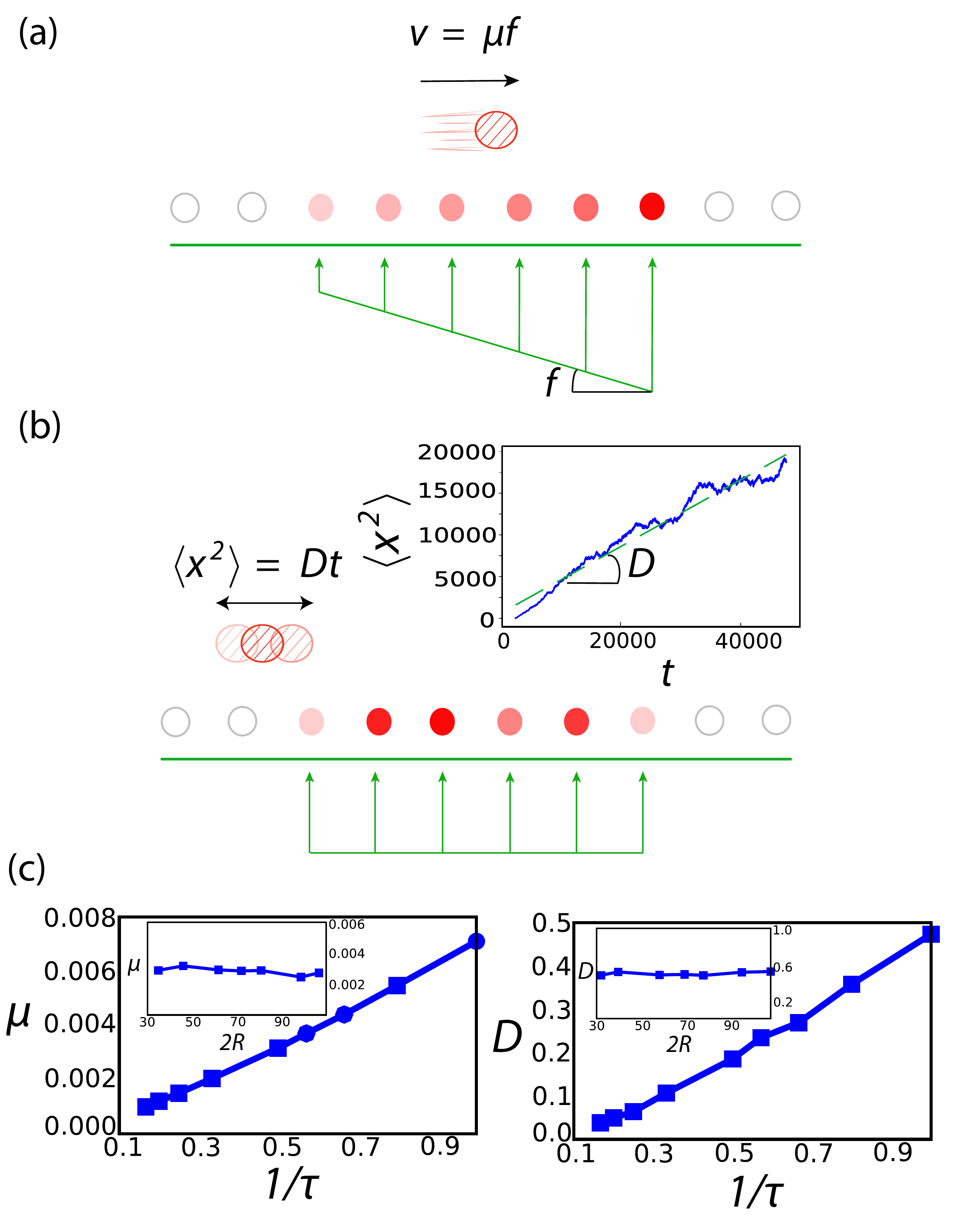}
\caption{(a) Schematics of the droplet being driven by a linear potential (ramp), illustrating the idea of mobility. Green lines are inputs, red dots are active neurons, the more transparent ones represent earlier time.
(b) Schematics of the droplet diffusing under an  input with no gradient, 
giving rise to diffusion. Inset is the plot of mean-squared distance vs time, clearly showing diffusive behavior. Note here we have changed the droplet c.o.m. position $\bar{x}$ as $x$ to avoid confusion with the mean-position.
(c) Comparison between mobility $\mu = \gamma^{-1}$ and diffusion coefficient $D$. Both $\mu$ and $D$ depend on blob size and $\tau$ in the same way, and thus $D$ is proportional to $\mu$. \label{fig:fluctuation-dissipation}}
\end{figure}

We use a finite ramp as the external driving, $I^{ext}_n = n$ with $n < n_{max}$, and $I^{ext}_n = 0$ otherwise (see Fig.\ref{fig:fluctuation-dissipation}(a)). We choose $n_{max}$ to be such that the  to the end of the ramp and still takes considerable time to relax to its steady-state position. We notice that for different slopes of the $I^{ext}_n$, the droplet have different velocities, and it is natural to define a mobility of the droplet, $\mu$, by $v = \mu f$, where $f$ is the slope of $I^{ext}_n$.
Next, we notice that on a single continuous attractor the droplet can diffuse because of internal noise in the neural network. Therefore, we can infer the diffusion coefficient $D$ of the droplet from $\langle x^2 \rangle = 2Dt$ for a collection of diffusive trajectories (see Fig.\ref{fig:fluctuation-dissipation}(b)), where we have used $x$ to denote the center of mass $\bar{x}$ for the droplet to avoid confusion.

In Fig.\ref{fig:fluctuation-dissipation}(c) we numerically verify that $\mu$ and $D$ depend on parameters $\tau$ and $R$ in the same way, i.e. $D$ and $\mu$ are both proportional to $1/\tau$ and independent of $R$. This suggest that $D \propto \mu$, if we call the proportionality constant to be $k_BT$, then we have a fluctuation-dissipation-like relation,
\begin{equation}
D = \mu k_B T.
\end{equation} 

\section{Space and time dependent external driving signals}


We consider the model of sensory input used in the main text: $I^{cup}(n) = d(w-|n|), n \in [-w,w]$, $I^{cup}(n) = 0$ otherwise.
We focus on time-dependent currents $I^{ext}_n(t) = I^{cup}(n - vt)$.
Such a drive was previously considered in \cite{Wu2005-sw}, albeit without time dependence. Throughout the paper, we refer to $w$ as the linear size of the drive, $d$ as the depth of the drive, and set the drive moving at a constant velocity $v$. From now on, we will go to the continuum limit and denote $I^{ext}_n(t) = I^{ext}(n,t) \equiv I^{ext}(x,t)$.

As an example, for $v=0$ (in this case, $\Delta x_v = \bar{x}$) we can write down the potential $V^{ext}$ for the external driving signal $I^{cup}(x) = d(w-|x|)$ by evaluating it at a stationary current profile $f(i_k^{\bar{x}}) = 1\; \text{if} \; |k-\bar{x}| \leq R, =0 \; \text{otherwise}$,
\begin{equation}
\label{eqn:Vext}
	V^{ext}(\bar{x}) = \begin{cases} 
		V_1(\bar{x}), & |\bar{x}| \leq R\\
		V_2(\bar{x}), & |\bar{x}| > R,
	\end{cases}
\end{equation}

where
\begin{equation}
	\begin{split}
		V_1(\bar{x}) &= -d\bigg [(R-\bar{x})(w-\frac{R-\bar{x}}{2}) + (R+\bar{x})(w-\frac{w+\bar{x}}{2})\bigg ]
		\\
		V_2(\bar{x}) &= -\frac{d}{2}(R+w-\bar{x})^2.
	\end{split}
\end{equation}

\begin{figure}
	\includegraphics[width=1\linewidth]{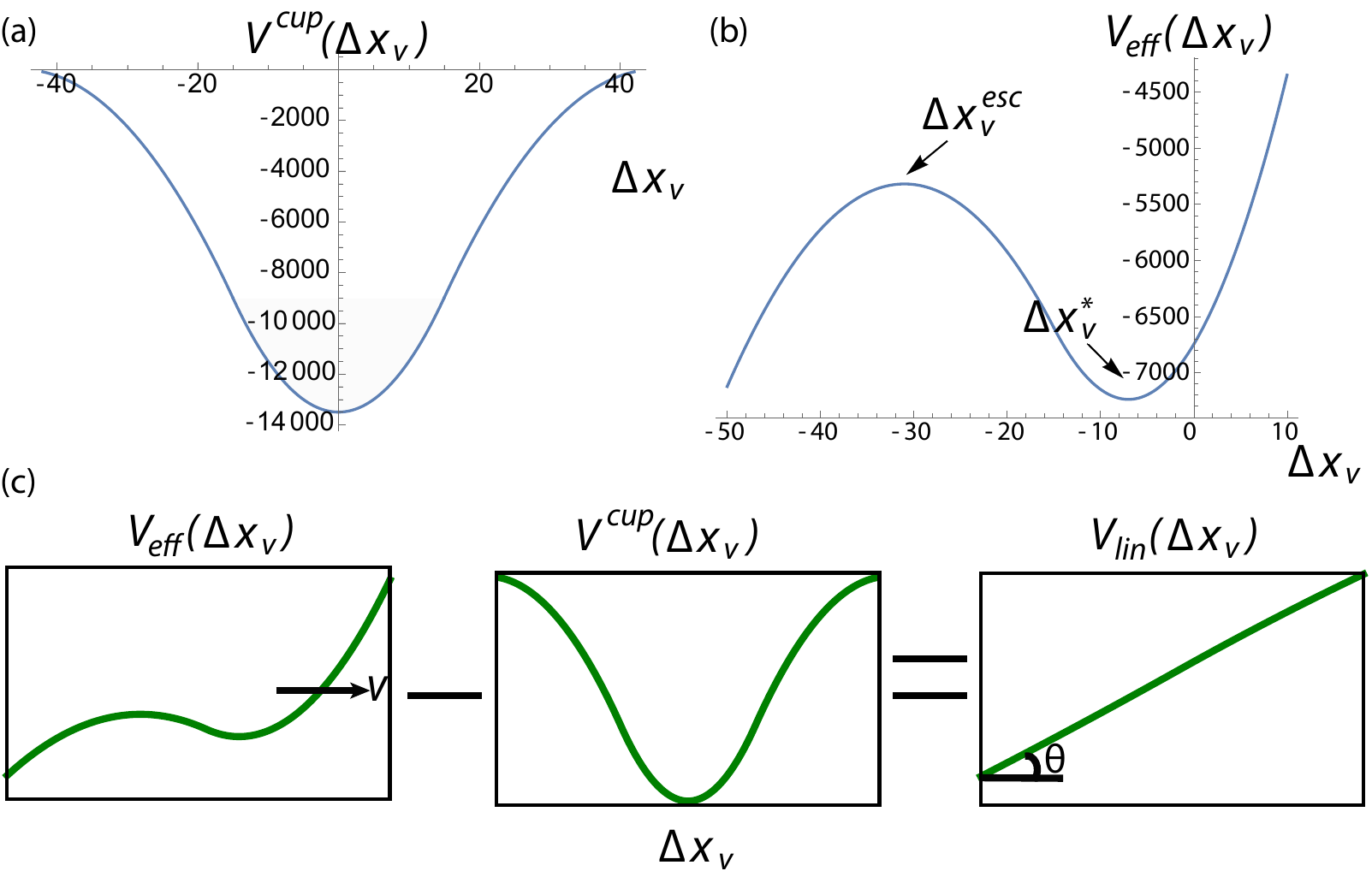}
	\caption{(a) $V^{ext}$ for external driving signal $I^{cup}(x,t)$ with $v=0$, plotted from Eqn.\eqref{eqn:Vext} with $d=20$, $R = 15$, $w = 30$. (b) Effective potential $V_{eff}$ experienced by the droplet for a moving cup-shaped external driving signal, plotted from Eqn.\eqref{eqn:Veff} with $d = 10$, $R= 15$, $w=30$, $\gamma v = 140$. (c) Schematic illustrating the idea of the equivalence principle (main text Eqn.\eqref{eqn:equivalance}). The difference between the effective potential, $V_{eff} \equiv -k_BT\log p(\Delta x_v)$, experienced by a moving droplet, and that of a stationary droplet, $V^{cup}$, is a linear potential, $V_{lin} = -F_v^{motion} \Delta x_v$. The slope $\theta$ of the linear potential $V_{lin} = - F_v^{motion}\Delta x_v$ is proportional to velocity as $F_v^{motion} = \gamma v$. \label{fig:Vext panel}}
\end{figure}

We plot $V^{ext}$ given by Eqn.\eqref{eqn:Vext} vs the c.o.m. position of droplet in Fig.\ref{fig:Vext panel}(a).

\subsection*{A thermal equivalence principle}
The equivalence principle we introduced in the main text allows us to compute the steady-state position and the effective new potential seen in the co-moving frame. Crucially, the fluctuations of the collective coordinate are described by the potential obtained through the equivalence principle. The principle correctly predicts both the mean (main text Eqn.\eqref{eqn:equivalance}) and the fluctuation (main text Eqn.\eqref{eqn:fluc lag}) of the lag $\Delta x_v$. Therefore, it is actually a statement about the equivalence of effective dynamics in the rest frame and in the co-moving frame. Specializing to the drive $I^{cup}(x,t)$, the equivalence principle predicts that the effective potential felt by the droplet (moving at constant velocity $v$) in the co-moving frame equals the effective potential in the stationary frame shifted by a linear potential, $V_{lin} = -F_v^{mot} \Delta x_v$, that accounts for the fictitious forces due to the change of coordinates (see Fig.\ref{fig:Vext panel}(c)).

Since we used \eqref{eqn:Vext} for the cup shape and the lag $\Delta x_v$ depends linearly on $v$, we expect that the slope of the linear potential $V_{lin}$ also depends linearly on $v$. Here the sign convention is chosen such that $V_{lin}<0$ corresponds to droplet moving to the right. 

\section{Speed limit for external driving signals}

In the following, we work in the co-moving frame with velocity $v$ at which the driving signal is moving. We denote the steady-state c.o.m. position in this frame to be $\Delta x^*_v$, and a generic position to be $\Delta x_v$.  

When $v>0$, the droplet will sit at a steady-state position $\Delta x_v^*<0$, equivalence principle says we should subtract a velocity-dependent linear potential $F^{mot}_v \Delta x_v = \gamma v \Delta x_v$ from $V^{ext}$ to account for the motion,
\begin{equation}
	\label{eqn:Veff}
	V_{eff}(\Delta x_v) = V^{cup}(\Delta x_v) - \gamma v \Delta x_v.
\end{equation}

We plot $V_{eff}$ vs $\Delta x_v$ in Fig.\ref{fig:Vext panel}(b). Notice that there are two extremal points of the potential, corresponding to the steady-state position, $\Delta x^*_v$, and the escape position, $\Delta x^{esc}_v$,

\begin{equation}
	\begin{split}
		\label{eqn:positions}
		\Delta x^*_v &= \gamma v/2d
		\\
		\Delta x^{esc}_v &= (dw-\gamma v + dR)/d.
	\end{split}
\end{equation}

We are now in position to derive $v_{crit}$ presented in the main text. We observe that as the driving velocity $v$ increases, $\Delta x^*_v$ and $\Delta x^{esc}_v$ will get closer to each other, and there will be a critical velocity such that the two coincide. 

By simply equating the expression for $x_{esc}$ and $x^*$ and solve for $v$, we found that
\begin{equation}
	\label{eqn:vcrit}
	v_{crit} = \frac{2d(w+R)}{3\gamma}.
\end{equation}

\subsection*{Steady-state droplet size}

Recall that the Lyapunov function of the neural network is given by \eqref{eqn:lyapunov},

\begin{equation}
\begin{split}
\mathcal{L}[\bar{x}]
&= \frac{1}{\tau}\sum_{k} \int_0^{f(i_k^{\bar{x}})}f^{-1}(x)dx\\
& + V_J(\bar{x})  + V^{ext}(\bar{x},t),
\end{split}
\end{equation}

Compared to the equation of motion \eqref{eqn:eom}, we see that the first term corresponds to the decay of neurons in the absence of interaction from neighbors (decay from 'on' state to 'off' state), and the second term corresponds to the interaction $J_{nk}$ term in the e.o.m, and the third term corresponds to the $I_n^{ext}$ in the e.o.m. Since we are interested in the steady-state droplet size, and thus only interested in the neurons that are 'on', the effect of the first term can be neglected (also note that $1/\tau \ll J_{ij}$, when using the Lyapunov function to compute steady-state properties, the first term can be ignored).

To obtain general results, we also account for long-ranged disordered connections $J^d_{ij}$ here. We assume $J^d_{ij}$ consists of random connections among all the neurons. We can approximate these random connections as random permutations of $J^0_{ij}$ and the full $J_{ij}$ is the sum over $M-1$ such permutations plus $J^0_{ij}$.

For the cup-shaped driving and its corresponding effective potential, Eqn.\eqref{eqn:Veff}, we are interested in the steady-state droplet size under this driving, so we first evaluate $V_{eff}$ at the steady-state position $\Delta x^*_v$ in Eqn.\eqref{eqn:positions}. To make the $R$-dependence explicit in the Lyapunov function, we evaluate $\mathcal{L}(\bar{x})$ under the 'rigid bump approximation' used in \cite{Hopfield2015-wt}, i.e., assuming $f(i_k^{\bar{x}}) = 1$ for $|k-\bar{x}| \leq R$, and $=0$ otherwise. 

We find that for $M-1$ sets of disorder interactions, the Lyapunov function is

\begin{equation}
	\begin{split}
		\label{eqn:blobenergy}
		\mathcal{L}[f(i_k^{\bar{x}})] &= J\bigg [ (\epsilon R^2 - (\epsilon + 2p)R + \frac{p(p+1)}{2} 
		\\
		&- pm(2R-p)^2    \bigg ] + \frac{(\gamma v)^2}{4d} + Rd(R-2w),
	\end{split}
\end{equation}
where we have defined the reduced disorder parameter $m = (M-1)/N$ and have used the equivalence principle in main text Eqn.\eqref{eqn:equivalance} to add an effective linear potential to take into account the motion of the droplet.

Next, we note that the steady-state droplet size corresponds to a local extremum of the Lyapunov function. Extremizing Eqn.\eqref{eqn:blobenergy} with respect to droplet radius $R$, we obtain the steady-state droplet radius as a function of the external driving parameters $d,w$, and the reduced disorder parameter $m$, 

\begin{equation}
	\label{eqn:blobsize}
	R(d,w,m) = \frac{2p-4p^2m+2wd/J+\epsilon}{2d/J - 8pm + 4\epsilon},
\end{equation}

where we observe that in the formula the only dimensionful parameters $d$ and $J$ appears together to ensure the overall result is dimensionless. Our result for $R$ reduces to $	R_0 = \frac{p}{2\epsilon} + \frac{1}{4}$ by setting $M=1$ and $d=w=0$.



\subsection*{Upper limit on external signal strength}

Here we present the calculation for maximal driving strength $I^{ext}$ 
beyond which the activity droplet will 'teleport' -- i.e., disappears at the original location and re-condense at the location of the drive, even if these two locations are widely separated. From now on, we refer to this maximal signal strength as the 'teleportation limit'. We can determine this limit by finding out the critical point where the energy barrier of breaking up the droplet at the original location is zero.

\begin{figure}
	\includegraphics[width=1\linewidth]{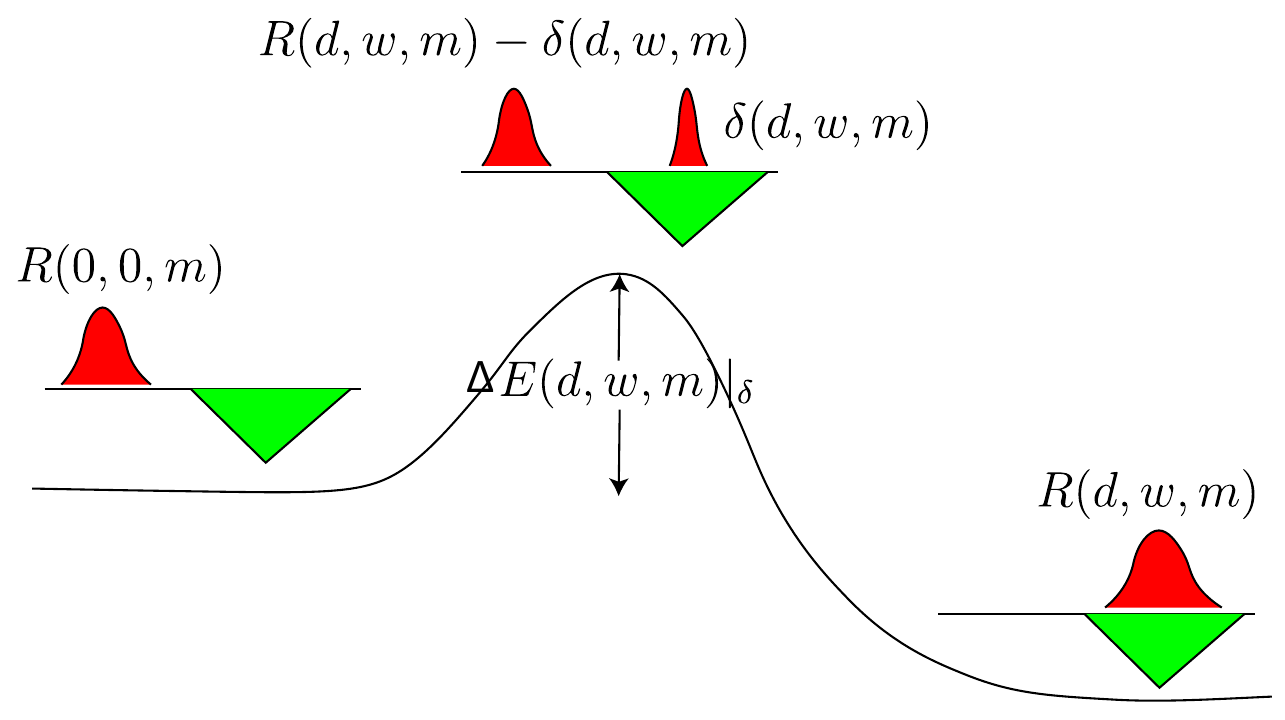}
	\caption{Schematics of three scenarios during a teleportation process. A initial configuration: the droplet is outside of the cup. A energetically unfavorable intermediate configuration that is penalize by $\Delta E$: the droplet breaks apart into two droplets, one outside the cup and one inside the cup; a final configuration with lowest energy: the droplet inside the cup grows to a full droplet while the droplet outside shrinks to zero size. Above each droplet is its corresponding radius $R$.\label{fig:teleportation}}
\end{figure}

For simplicity, we assume that initially the cup-shaped driving signal is some distance $x_0$ from the droplet, and not moving (the moving case can be solved in exactly the same way by using equivalence principle and going to the co-moving frame of the droplet). We consider the following three scenarios during the teleportation process: $(1)$ the initial configuration: the droplet have not yet teleported, and stays at the original location with radius $R(0,0,m)$; $(2)$ the intermediate configuration: where the activity is no longer contiguous, giving a droplet with radius $\delta(d,w,m)$ at the center of the cup, and another droplet with radius $R(d,w,m)-\delta(d,w,m)$ at the original location (when teleportation happens, the total firing neurons changes from $R(0,0,m)$ to $R(d,w,m)$); $(3)$ the final configuration: the droplet has successfully teleported to the center of the cup, with radius $R(d,w,m)$. The three scenarios are depicted schematically in Fig.\ref{fig:teleportation}.

The global minimum of the Lyapunov function corresponds to scenario $(3), $ However, there is an energy barrier between the initial configuration $(1)$ and final configuration $(3)$, corresponding to the $V_{eff}$ difference between initial configuration $(1)$ and intermediate configuration $(2)$. We would like to find the critical split size $\delta_c(d,w,m)$ that maximize the difference in $V_{eff}$, which corresponds to the largest energy barrier the network has to overcome in order to teleporte from $(1)$ to $(3)$. For the purpose of derivation, in the following we would like to rename $\mathcal{L}[f(i_k^m)]$ in Eqn.\eqref{eqn:blobenergy} as $E_0(d,w,m)\rvert_{R(d,w,m)}$ to emphasize its dependence on the external driving parameters and disordered interactions. The subscript $0$ stands for the default one-droplet configuration, and it is understood that $E_0(d,w,m)$ is evaluated at the network configuration of a single droplet at location $m$ with radius $R(d,w,m)$.

The energy for $(1)$ is simply $E_0(0,0,m)$, and the energy for $(3)$ is $E_0(d,w,m)$. However, the energy for $(2)$ is not just the sum of $E_0$ from the two droplets. Due to global inhibitions presented in the network, when there are two droplets, there will be an extra interaction term, when we evaluate the Lyapunov function with respect to this configuration. The interaction energy between two droplets in Fig.\ref{fig:teleportation} is

\begin{equation}
	E_{int}(m)\rvert_{R,\delta} = 4JR\delta(\epsilon - 2pm).
\end{equation}

Therefore, the energy barrier for split size $\delta$ is
\begin{equation}
	\begin{split}
		\Delta E&(d,w,m)\rvert_{\delta}  \\
		&= E_0(0,0,m)\rvert_{R(d,w,m)-\delta} + E_0(d,w,m)\rvert_{\delta}
		\\
		&+ E_{int}(m)\rvert_{R(d,w,m),\delta} - E_0(0,0,m)\rvert_{R(0,0,m)}.
	\end{split}
\end{equation}

Therefore, maximizing $\Delta E$ with respect to $\delta$, we find
\begin{equation}
	\begin{split}
		\delta_c &= \frac{dw}{d-8Jpm + 4J\epsilon}
	\end{split}
\end{equation}

Now we have obtained the maximum energy barrier during a teleportation process, $\Delta E\rvert_{\delta_c}$. A spontaneous teleportation will occur if $\Delta E\rvert_{\delta_c} \leq 0$, and this in turn gives a upper bound on external driving signal strength $d \leq d_{max}$ one can have without any teleportation spontaneous occurring. 

\begin{figure}
	\includegraphics[width=1\linewidth]{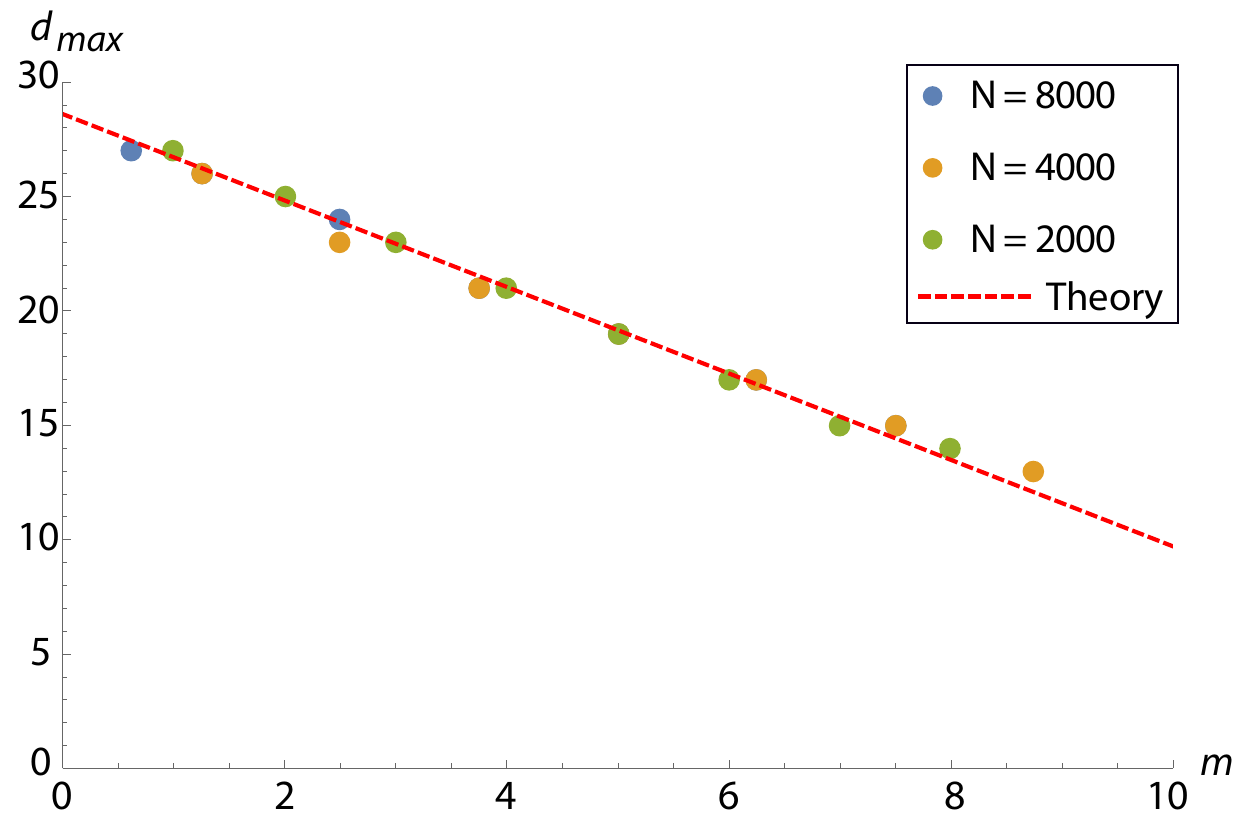}
	\caption{Teleportation depth $d_{max}$ plotted against disorder parameter $m$. The dots are data obtained from simulations for different $N$ but with $p=10$, $\epsilon=0.35$, $\tau=1$, $J=100$, and $w=30$ held fixed. The dotted line is the theoretical curve plotted from solving $\Delta E(d_c,w,m)\rvert \delta_c = 0$ for $d_{c}$ numerically. \label{fig:dmax}}
\end{figure}

We plot the numerical solution of $d_{max}$ obtained from solving $\Delta E(d_c,w,m)\rvert \delta_c = 0$, compared with results obtained from simulation in Fig.\ref{fig:dmax}, and find perfect agreement. 


We also obtain an approximate solution by observing that the only relevant scale for that the critical split size $\delta_c$ is the radius of the droplet, $R$. We set $\delta_c = cR$ for some constant $0 \leq c \leq 1$. In general, $c$ can depend on dimensionless parameters like $p$ and $\epsilon$. Empirically we found the constant to be about 0.29 in our simulation. 

The droplet radius $R$ is a function of $d,w,m$ as we see in Eqn.\eqref{eqn:blobsize}, but to first order approximation we can set $R$ = $R^*$ for some steady-state radius $R^*$. Then we can solve

\begin{equation}
	\label{eqn:dmax}
	d_{max}(M) = \frac{4J(\epsilon-2pm)}{w/cR^* -1}.
\end{equation}

Note that the denominator is positive because $w > R$ and $0 \leq c \leq 1$. The simulation result also confirms that the critical split size $\delta_c$ stays approximately constant. 
We have checked that the dependence on parameters $J,w, m$ in Eqn.\eqref{eqn:dmax} agrees with the numerical solution obtained from solving $E_{bar}(d_c,w,m)\rvert \delta_c = 0$, up to the undetermined constant $c$.

\subsection*{Speed limit on external driving}
\label{app:v_fund}
Recall that given a certain signal strength $d$, there is an upper bound on how fast the driving can be, Eqn.\eqref{eqn:vcrit}. Then in particular, for $d_{max}$, we obtain an upper bound on how fast external signal can drive the network,
\begin{equation}
	v_{max} = \frac{8J(w+R^*)(\epsilon-2pm)}{3\gamma(w/cR^* -1)}.
\end{equation}

For $w \gg R^*$, we can approximate
\begin{equation}
	v_{max} \approx \frac{16JcR^*(\epsilon/2-pm)}{3\gamma},
\end{equation}

In the absence of disorder, $m = 0$, the maximum velocity is bounded by

\begin{equation}
	\label{eqn:vbound_1}
	v_{max} \leq \frac{8c}{3}\frac{\epsilon JR^*}{\gamma} \leq \frac{8c}{3}\frac{\epsilon JR_{max}}{\gamma}.
\end{equation}

Recall that in Eqn.\eqref{eqn:dmax}, we have
%
%
\begin{equation}
	\begin{split}
		R(d,w\gg R,0) &\leq R(d_{max}, w\gg R, 0)
		\\
		&= \frac{p}{2\epsilon} + \frac{1}{4} + 2cR^* + \mathcal{O}(\frac{R}{w})
		\\
		& \lessapprox \frac{p}{2\epsilon} + 2cR_{max},
	\end{split}
\end{equation}

where in the second line we have used \eqref{eqn:blobsize} for $d=d_{max}$, $m=0$, and $w \gg R$. 
Upon rearranging, we have

\begin{equation}
R_{max} \lessapprox \frac{1}{1-2c}\frac{p}{2\epsilon}.
\end{equation}

Plugging in Eqn.\eqref{eqn:vbound_1}, we have

\begin{equation}
	v_{max} \leq \frac{8c}{3}\frac{\epsilon JR_{max}}{\gamma} \lessapprox \frac{8}{3(c^{-1}-2)} \frac{Jp}{\gamma}.
\end{equation}

Therefore, we have obtained an fundamental limit on how fast the droplet can move under the influence of external signal, namely,
\begin{equation}
	v_{fund} = \kappa Jp\gamma^{-1},
\end{equation}

where $\kappa = 8/3(c^{-1}-2)$ is a dimensionless $\mathcal{O}(1)$ number.

\section{Path integration and velocity input}

\label{Aij}
Place cell networks \cite{Ocko2018-gv} and head direction networks \cite{Kim2017-zs} are known to receive information both about velocity and landmark information. Velocity input can be modeled by adding an anti-symmetric part $A_{ij}$ to the connectivity matrix $J_{ij}$, which effectively 'tilts' the continuous attractor. 

Consider now 
\begin{equation}
J_{ij} = J^0_{ij} + J^d_{ij} + A^0_{ij}, 
\end{equation}
where $A^0_{ij} = A$, if $0< i-j \leq p$; $-A$, if $0< j-i \leq p$; and $0$ otherwise.  

\begin{figure}
	\includegraphics[width=0.6\linewidth]{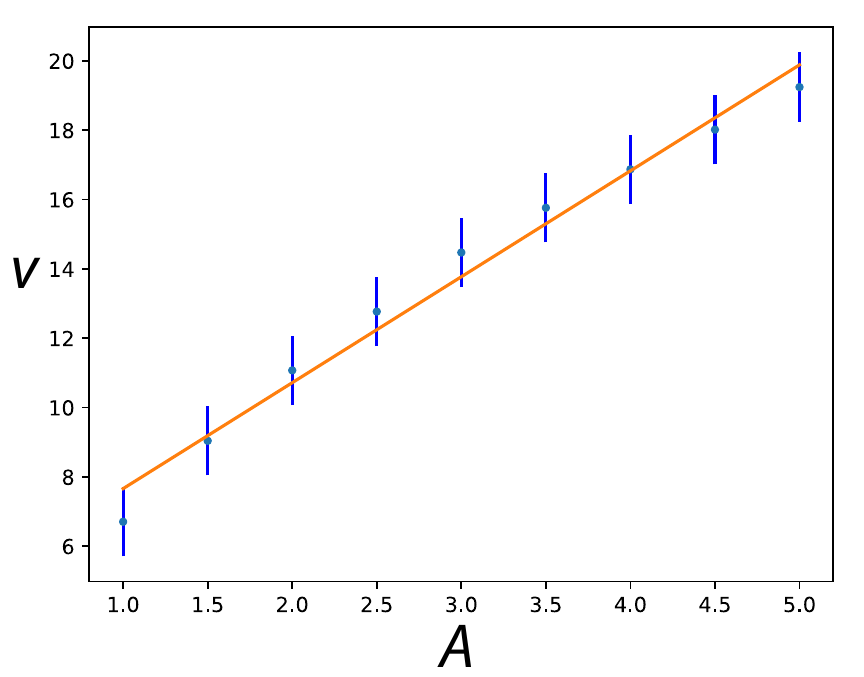}
	\caption{Velocity of droplet $v$ plotted against the size $A$ of the anti-symmetric matrix. We hold all other parameters fixed with the value same as in Fig.\ref{fig:dmax}. \label{fig:vA}}
\end{figure}

The anti-symmetric part $A^0_{ij}$ will provide a velocity $v$ that is proportional to the size $A$ of $A^0_{ij}$ for the droplet (See Fig.\ref{fig:vA}). In the presence of disorder, we can simply go to the co-moving frame of velocity $v$ and the droplet experiences an extra disorder-induced noise $\eta_A$ in addition to the disorder induced temperature $T_d$. 

\begin{figure}
	\includegraphics[width=1\linewidth]{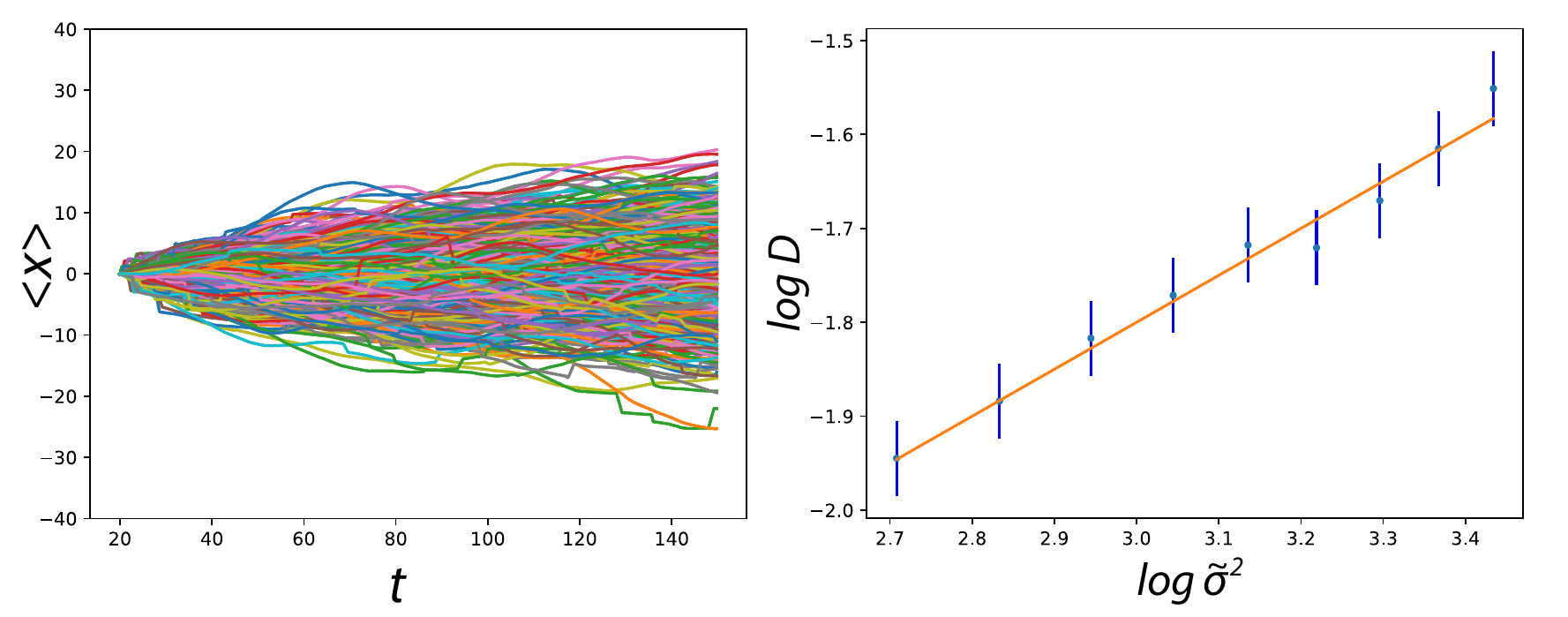}
	\caption{\textbf{Left}: At fixed $A=5$, a collection of 500 diffusive trajectories in the co-moving frame at velocity $v$, where $v$ is taken to be the average velocity of all the trajectories. We can infer the diffusion coefficient $D$ from the variance of these trajectories as Var$(x)= 2Dt$. \textbf{Right}: log$D$ plotted against log$\tilde{\sigma}^2$. The straight line has slope $1/2$, corresponding to $D \propto \tilde{\sigma}$. \label{fig:asym}}
\end{figure}

We found that $\langle \eta_{A}(t)\eta_{A}(0)\rangle \propto \tilde{\sigma}\delta(t)$ (See Fig.\ref{fig:asym}), where $\tilde{\sigma}^2$ is the average number of disordered connection per neuron in units of $2p$.

Therefore, all our results in the main text applies to the case when both the external drive $I^{ext}(x,t)$ and the anti-symmetric part $A^0_{ij}$ exists. Specifically, we can just replace the velocity $v$ used in the main text as the sum of the two velocities corresponding to $I^{ext}(x,t)$ and $A^0_{ij}$.

\section{Quenched Disorder - driving and disorder-induced temperature}
\label{Tbumpy}

\subsection{Disordered connections and disordered forces}

From now on, we start to include disorder connections $J^d_{ij}$ in additional to ordered connections $J^0_{ij}$ that corresponds to the nearest $p$-neighbor interactions. We assume $J^d_{ij}$ consists of random connections among all the neurons. These random connections can be approximated as random permutations of $J^0_{ij}$, such that the full $J_{ij}$ is the sum over $M-1$ such permutations plus $J^0_{ij}$. 

We `clip' the $J_{ij}$ matrix according to the following rule for each entry when summing over $J^0_{ij}$ and $J^d_{ij}$,
\begin{equation}
\label{eqn:clip}
\begin{split}
J&(1-\epsilon) + J(1-\epsilon) \to J(1-\epsilon) \\
J&(1-\epsilon) + J(-\epsilon) \to J(1-\epsilon)  \\
J&(-\epsilon) + J(-\epsilon) \to J(-\epsilon).
\end{split}
\end{equation}
Therefore, adding more disorder connections to $J_{ij}$ amounts to changing the inhibitory $-J\epsilon$ entries to the exitory $J(1-\epsilon)$. 

We would like to characterize the effect of disorder on the system. Under the decomposition $J_{ij} = J^0_{ij} + J^d_{ij}$, we can define a (quenched) disorder potential 
\begin{equation}
V^{d}(\bar{x}) \equiv V^{d}[f(i_k^{\bar{x}})] = -\frac{1}{2} \sum_{nk}J^{d}_{nk}f(i_k^{\bar{x}})f(i_n^{\bar{x}}),
\end{equation}

that captures all the disorder effects on the network.

Its corresponding disorder-induced force is then given by
\begin{equation}
\label{eqn:Fd}
 F^d(\bar{x}) = -\partial_{\bar{x}} V^{d}(\bar{x}).
\end{equation}

\subsection{Variance of disorder forces}

We compute the distribution of $V^d(\bar{x})$ using a combinatorial argument as follows. 

Under the rigid droplet approximation, calculating $V^d(\bar{x})$ amounts to summing all the entries within a $R$-by-$R$ diagonal block sub-matrix $J^{(\bar{x})}_{ij}$ within the full synaptic matrix $J_{ij}$ (recall that $V^d(\bar{x}) \propto \sum_{nk}f(i_n^{(\bar{x})})J_{nk}f(i_k^{(\bar{x})})$). Each set of disorder connection is a random permutation of $J^0_{ij}$, and thus has the same number of excitatory entries as $J^0_{ij}$, namely $2pN$. Since the inhibitory connections do not play a role in the summation by the virtue of \eqref{eqn:clip}, it suffices to only consider the effect of adding excitatory connections in $J^d_{ij}$ to $J^0_{ij}$.
 
There are $M-1$ sets of disordered connections in $J^d_{ij}$, and each has $2pN$ excitatory connections. Now suppose we add these $2pN(M-1)$ excitatory connections one by one to $J^0_{ij}$. Each time an excitatory entry is added to an entry $y$ in the $R$-by-$R$ block $J^{(\bar{x})}_{ij}$, there are two possible situations depending on the value of $y$ before addition: if $y = J(1-\epsilon)$ (excitatory), the addition of an excitatory connection does not change the value of $y$ because of the clipping rule in \eqref{eqn:clip}; if $y = -J\epsilon$ (inhibitory), the addition of an excitatory connection to $y$ changes $y$ to $J(1-\epsilon)$. In the latter case the value of $V^d(\bar{x})$ is changed because the summation of entries within $J^{(\bar{x})}_{ij}$ has changed, while in the former case $V^d(\bar{x})$ stays the same. (Note that if the excitatory connection is added outside $J^{(\bar{x})}_{ij}$, it does not change $V^d(\bar{x})$ and thus can be neglected.)

We have in total $2pN(M-1)$ excitatory connections to be added, and in total $(2R-p)^2$ potential inhibitory connections in the $R$-by-$R$ block $J^{(\bar{x})}_{ij}$ to be `flipped' to an excitatory connection. We are interested in, after adding all the $2pN(M-1)$ excitatory connections how many inhibitory connections are changed to excitatory connections, and the corresponding change in $V^d(\bar{x})$. 


We can get an approximate solution if we assume that the probability of flipping an inhibitory connection does not change after subsequent addition of excitatory connections, and stays constant throughout the addition of all the $2pN(M-1)$ excitatory connections. This requires $2pN(M-1) \ll N^2$, i.e., $M \ll N$, which is a reasonable assumption since the capacity can not be $\mathcal{O}(N)$. 

For a single addition of exitatory connection, the probability of successfully flipping an inhibitory connection within $J^{(\bar{x})}_{ij}$ is proportional to the fraction of the inhibitory connections within $J^{(\bar{x})}_{ij}$ over the total number of entires in $J^0_{ij}$, 
\begin{equation}
q(\text{flip}) = \frac{(2R-p)^2}{N^2}.
\end{equation}

So the probability of getting $n$ inhibitory connections flipped is
\begin{equation}
P(n) = {2pN(M-1)\choose n} q^n (1-q)^{2pN(M-1)-n}.
\end{equation}

In other words, the distribution of flipping $n$ inhibitory connections to excitatory connections after adding $J^d_{ij}$ to $J^0_{ij}$ obeys $n \sim B(2pN(M-1),q)$. The mean is then
\begin{equation}
\begin{split}
\langle n \rangle &= 2pN(M-1)q = 2p(2R-p)^2 \bigg(\frac{M-1}{N}\bigg) \\
&= (2R-p)^2 2pm,
\end{split}
\end{equation}

where we have defined the reduced disorder parameter $m \equiv (M-1)/N$. The variance is 

\begin{equation}
\begin{split}
\langle n^2 \rangle &= 2pN(M-1)q(1-q) \\
&= 2pN(M-1) \frac{(2R-p)^2}{N^2} \bigg(1-\frac{(2R-p)^2}{N^2}\bigg) \\
&\approx (2R-p)^2 2pm,
\end{split}
\end{equation}

where in the last line we have used $N \gg 2R-p$.

Since changing $n$ inhibitory connections to $n$ exitory connections amounts to changing $V^d(\bar{x})$ by $-1/2 (J(1-\epsilon) - J(-\epsilon)) = -J/2$, we have

\begin{equation}
\label{eqn:Vd}
\text{Var}(V^d(\bar{x})) \equiv \sigma^2 = J^2(R-p/2)^2 pm.
\end{equation}

\subsection{Disorder temperature from disorder-induced force}

We focus on the case where $I^{ext}_n$ gives rise to a constant velocity $v$ for the droplet (as in the main text). In the co-moving frame, the disorder-induced force $F^d(\bar{x})$ acts on the c.o.m. like random kicks with correlation within the droplet size. For fast enough velocity those random kicks are sufficiently de-correlated and become a white noise at temperature $T_d$. 

To extract this disorder-induced temperature $T_d$, we consider the autocorrelation of $F^{d}[\bar{x}(t)]$ between two different c.o.m. location $\bar{x}(t)$ and $\bar{x}'(t')$ (and thus different times $t$ and $t'$),

\begin{equation}
C(t,t') \equiv \langle F^{d}[\bar{x}(t)] F^{d}[\bar{x}(t')]\rangle,
\end{equation}

where the expectation value is averaging over different realizations of the quenched disorder. 

Using \eqref{eqn:Fd}, we have
\begin{eqnarray}
C(t,t') &=  \langle \partial_{\bar{x}} V^d(\bar{x}) \partial_{\bar{x}'}V^d(\bar{x}') \rangle \\
&=\partial_{\bar{x}}\partial_{\bar{x}'} \langle  V^d(\bar{x}) V^d(\bar{x}') \rangle.
\end{eqnarray}

Within time $t-t'$, if the droplet moves a distance less than its size $2R$, then $V^{d}$ computed at $t$ and $t'$ will be correlated because $f(i_k^{\bar{x}})$ and $f(i_k^{\bar{x'}})$ have non-zero overlap. Therefore, we expect the autocorrelation function $\langle  V^d(\bar{x}) V^d(\bar{x}') \rangle$ behaves like the 1-$d$ Ising model with finite correlation length $\xi = 2R$ (up to a prefactor to be fixed later), 

\begin{equation}
\langle  V^d(\bar{x}) V^d(\bar{x}') \rangle \sim \exp (-\frac{|\bar{x}-\bar{x}'|}{\xi}).
\end{equation}



Hence, $C(t,t') \sim \exp \bigg(-\frac{|\bar{x}-\bar{x}'|}{\xi}\bigg)$. Now going to the co-moving frame, we can write the c.o.m. location as before, $\Delta x_v = \bar{x} - vt$, so the autocorrelation function becomes


\begin{equation}
\begin{split}
C(t,t') &\sim  \exp \bigg(-\frac{|(\Delta x_v + vt) - (\Delta x'_v +vt')|}{\xi}\bigg) \\
&=  \exp \bigg(-\frac{|v(t-t') + (\Delta x_v - \Delta x'_v)|}{\xi}\bigg) \\
&\approx \exp \bigg(-\frac{v|t-t'|}{\xi}\bigg),
\end{split}
\end{equation}

where in the last line we have used that the droplet moves much faster in the stationary frame than the c.o.m. position fluctuates in the co-moving frame, so $v(t-t') \gg \Delta x_v - \Delta x'_v$. 

Now let us define the correlation time to be $\tau_{cor} = \xi/v = 2R/v$. Then

\begin{equation}
C(t,t') \sim \exp \bigg(-\frac{|t-t'|}{\tau_{cor}}\bigg).
\end{equation}

For $T \equiv |t-t'|\gg \tau_{cor}$, we want to consider the limiting behavior of $C(t,t')$ under an integral. Note that 

\begin{equation}
\begin{split}
&\int_{0}^{T} dt  \int_{0}^{T} dt' \exp \bigg(-\frac{|t-t'|}{\tau_{cor}}\bigg) \\
&= \tau_{cor}[2(T-\tau_{cor})+2\tau_{cor}e^{-T/\tau_{cor}}] \\
&\approx 2\tau_{cor}T \;\;\;\;\quad (\mbox{if } T \gg \tau_{cor}).
\end{split}
\end{equation}

Therefore, we have for $T \gg \tau_{cor}$, 

\begin{equation}
\begin{split}
&\int_{0}^{T} dt  \int_{0}^{T} dt' \exp \bigg(-\frac{|t-t'|}{\tau_{cor}}\bigg) \\
&= 2\tau_{cor} \int_{0}^{T} dt  \int_{0}^{T} dt' \delta (t-t').
\end{split}
\end{equation}

So we can write
\begin{equation}
\exp \bigg(-\frac{|t-t'|}{\tau_{cor}}\bigg) \to 2\tau_{cor} \delta(t-t'),
\end{equation}

and it is understood that this holds in the integral sense. Therefore, for $T \gg \tau_{cor}$, we expect $F^{d}(x)$ to act like uncorrelated white noise and we can write,

\begin{equation}
C(t,t') = T_d \delta(t-t') \propto \tau_{cor} \delta(t-t')
\end{equation}
where $T_d$ is a measure of this disorder-induced white noise. 

\begin{figure}
	\includegraphics[width=1\linewidth]{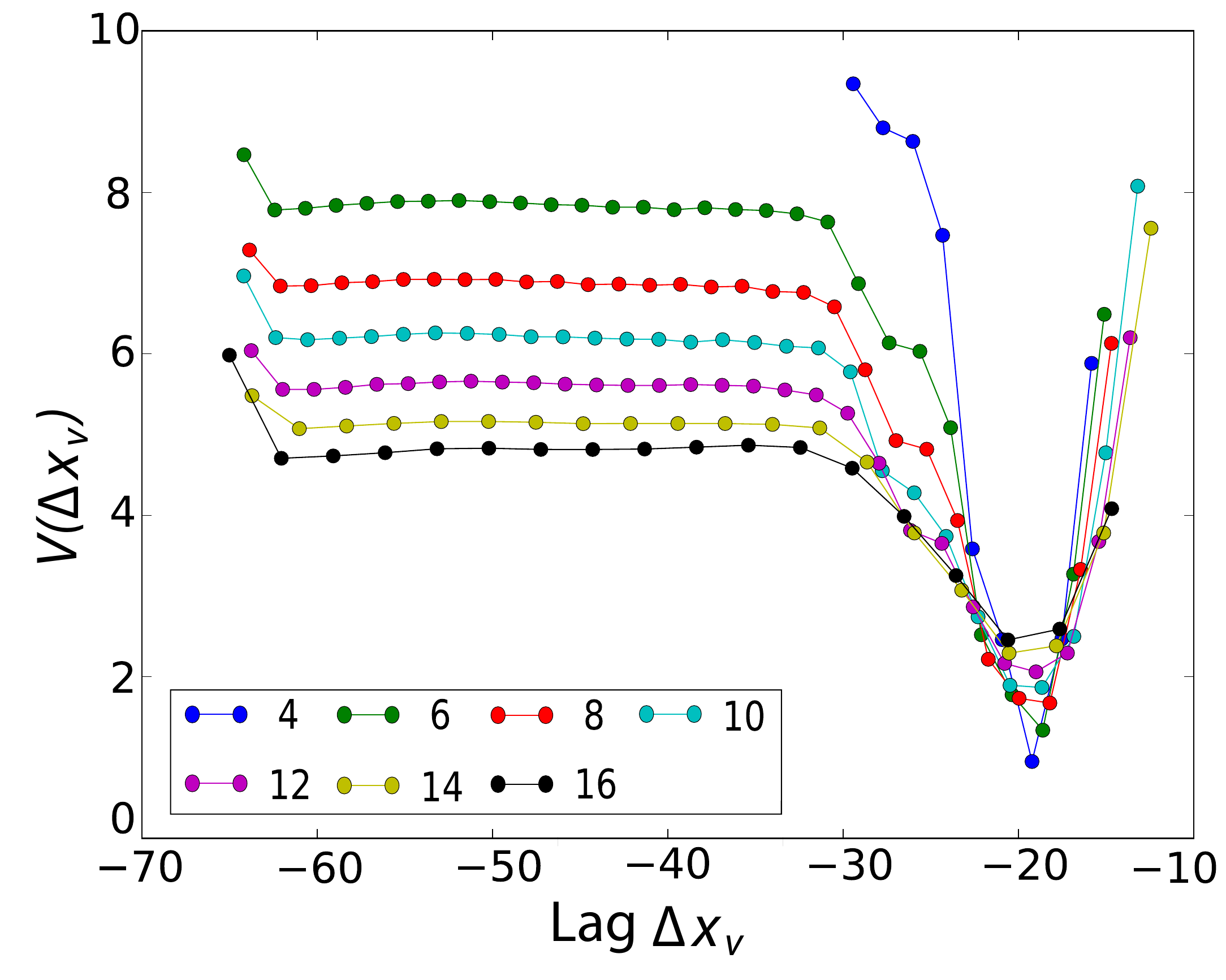}
	\caption{Uncollapsed data for the occupancies $- \log p(\Delta x_v)$ for different amounts of long ranged disordered connections.  Parameters same as in main text Fig.\ref{fig:Tbumpy} (see the last section of SI for further details). \label{fig:T_bumpy_uncollapsed}}
\end{figure}

To deduce the form of disorder temperature $T_d$, we present the uncollapsed occupancies $- \log p(\Delta x_v) = V(\Delta x_v)/k_B T_d$ (described in the caption of main text Fig.\ref{fig:Tbumpy}) in Fig.\ref{fig:T_bumpy_uncollapsed}. Compare with main text Fig.\ref{fig:Tbumpy}, we can see that $T_d$ successfully captures the effect of disorder on the statistics of the emergent droplet if, 

\begin{equation}
\label{eqn:Td}
T_d = \tilde{k} \tau_{cor} \sigma,
\end{equation}
where $\sigma$ is given in \eqref{eqn:Vd} and $\tilde{k}$ is a fitting constant.

\section{Derivation of the memory capacity for driven place cell network}
\label{pret}
In this section, we derive the memory capacity for driven place cell network described in the last section of the paper, namely, main text Eqn.\eqref{eqn:capacity}.

Our continuous attractor network can be applied to study the place cell network. We assume a 1-dimensional physical region of length $L$. We study a network with $N$ place cell neurons and assume each neuron has a place field of size $d = 2 p L/N$ that covers the region $[0,L]$ as a regular tiling. The $N$ neurons are assumed to interact as in the leaky integrate-and-fire model of neurons. The external driving currents $I^{ext}(x,t)$ can model sensory input when the mouse is physically in a region covered by place fields of neurons $i, i+1,\ldots, i+z$, currents $I^{ext}_{i}$ through $I^{ext}_{i+z}$ can be expected to be high compared to all other currents $I^{ext}_j$, which corresponds to the cup-shape drive we used throughout the main text.

It has been shown in past work that the collective coordinate in the continuous attractor survives to multiple environments provided the number of stored memories $m < m_c$ is below the capacity $m_c$ of the network. Under capacity, the neural activity droplet is multistable; that is, neural activity forms a stable contiguous droplet as seen in the place field arrangement corresponding to any one of the $m$ environments. Note that such a contiguous droplet will not appear contiguous in the place field arrangement of any other environment. Capacity was shown to scale as $m_c = \alpha(p/N, R) N$ where $\alpha$ is an $O(1)$ number that depends on the size of the droplet $R$ and the range of interactions $p$.  However, this capacity is about the intrinsic stability of droplet and does not consider the effect of rapid driving forces.


When the droplet escapes from the driving signal, it has to overcome certain energy barrier. This is the difference in $V_{eff}$ between the two extremal points $\Delta x^*_v$ and $\Delta x^{esc}_v$. Therefore, we define the barrier energy to be $\Delta E = V_{eff}(x^{esc}_v) - V_{eff}(\Delta x^*_v)$, and we evaluate it using Eqn.\eqref{eqn:Veff} and Eqn.\eqref{eqn:positions},
\begin{equation}
\label{eqn:deltaE}
\Delta E(v,d) = \frac{(4dw-3\gamma v -2dR)(-\gamma v+2dR)}{4d}.
\end{equation}

Note this is the result we used in main text Eqn.\eqref{eqn:capacity}.

As in the main text, the escape rate $r$ is given by the Arrhenius law,

\begin{equation}
r \sim \exp(-\frac{\Delta E(v,d)}{k_B T_{d}}).
\end{equation}

The total period of time of an external drive moving the droplet across a distance $L$ ($L\leq N$, but without loss of generality, we can set $L = N$) is $T = L/v$. We can imagine chopping $T$ into infinitesimal intervals $\Delta t$ st the probability of successfully moving the droplet across $L$ without escaping is,

\begin{equation}
\begin{split}
P_{retrieval} &= \lim_{\Delta t \to 0} (1-r \Delta t)^{\frac{T}{\Delta t}} 
\\
&= e^{-rT} = e^{-rN/v}
\\
&= \exp(-\frac{N}{v}e^{-\Delta E(v,d)/k_B T_{d}}).
\\
\end{split}
\end{equation}

$T_{d}$ is given by Eqn.\eqref{eqn:Td}

\begin{equation}
\begin{split}
T_d &=  \frac{2\tilde{k}R J(R-p/2)\sqrt{pm}}{v} \\
&\equiv k\sqrt{m}v^{-1},
\end{split}
\end{equation}

where in the last step we have absorbed all the constants (assuming $R$ is constant over different $m$'s) into the definition of $k$. 
Now we want to find the scaling behavior of $m$ s.t. in the thermodynamic limit ($N\to \infty$), $P_{retrieval}$ becomes a Heaviside step function $\Theta (m_c-m)$ at some critical memory $m_c$. With the aid of some hindsight, we try

\begin{equation}
m = \frac{\alpha^2}{(\log N) ^2},
\end{equation}

then in the thermodynamic limit,

\begin{equation}
\begin{split}
\lim_{N\to \infty} P_{retrieval} &= \lim_{N\to \infty} \exp(-\frac{N}{v}e^{-\log N v\Delta E(v,d)/\alpha k_B k})
\\
&= \lim_{N\to \infty} \exp (-\frac{N}{v} N^{-v\Delta E(v,d)/\alpha k_B k})
\\
&= \lim_{N\to \infty} \exp(-\frac{1}{v}N^{1-v\Delta E(v,d)/\alpha k_B k})
\\
&= \begin{cases} 
      1, & \alpha < v \Delta E(v,d)/k_B k \\
      0, & \alpha > v \Delta E(v,d)/k_B k
\end{cases}
\end{split}
\end{equation}

Therefore, we have arrive at the expression for capacity $m_c$, or in terms of $M = m_c N +1 \approx m_c N (N\gg 1)$,

\begin{equation}
M_c = \bigg[\frac{v\Delta E(v,d)}{k_B k}\bigg]^2 \frac{N}{(\log N)^2},
\end{equation}

or

\begin{equation}
M_c \sim \bigg[v\Delta E(v,d)\bigg]^2 \frac{N}{(\log N)^2}.
\end{equation}

\subsection*{Numerics of the place cell network simulations}
In this section, we explain our simulations in main text Fig.\ref{fig:capacity} in detail. 

Recall that we only determine the Arrhenius-like escape rate $r$ up to an overall constant, we can absorb it into the definition of $\Delta E(v,d)$ (given by Eqn.\eqref{eqn:deltaE}) as an additive constant $a$,

\begin{equation}
r = \exp\bigg \{{-\frac{\Delta E(v,d)+a}{k_B k v\sqrt{(M-1)/N}}}\bigg \}.
\end{equation}

Then the theoretical curves corresponds to
\begin{equation}
\label{eqn:pretrieval}
P_{retrieval} = e^{-Nr/v}
\end{equation}

\begin{figure}
	\includegraphics[width=1\linewidth]{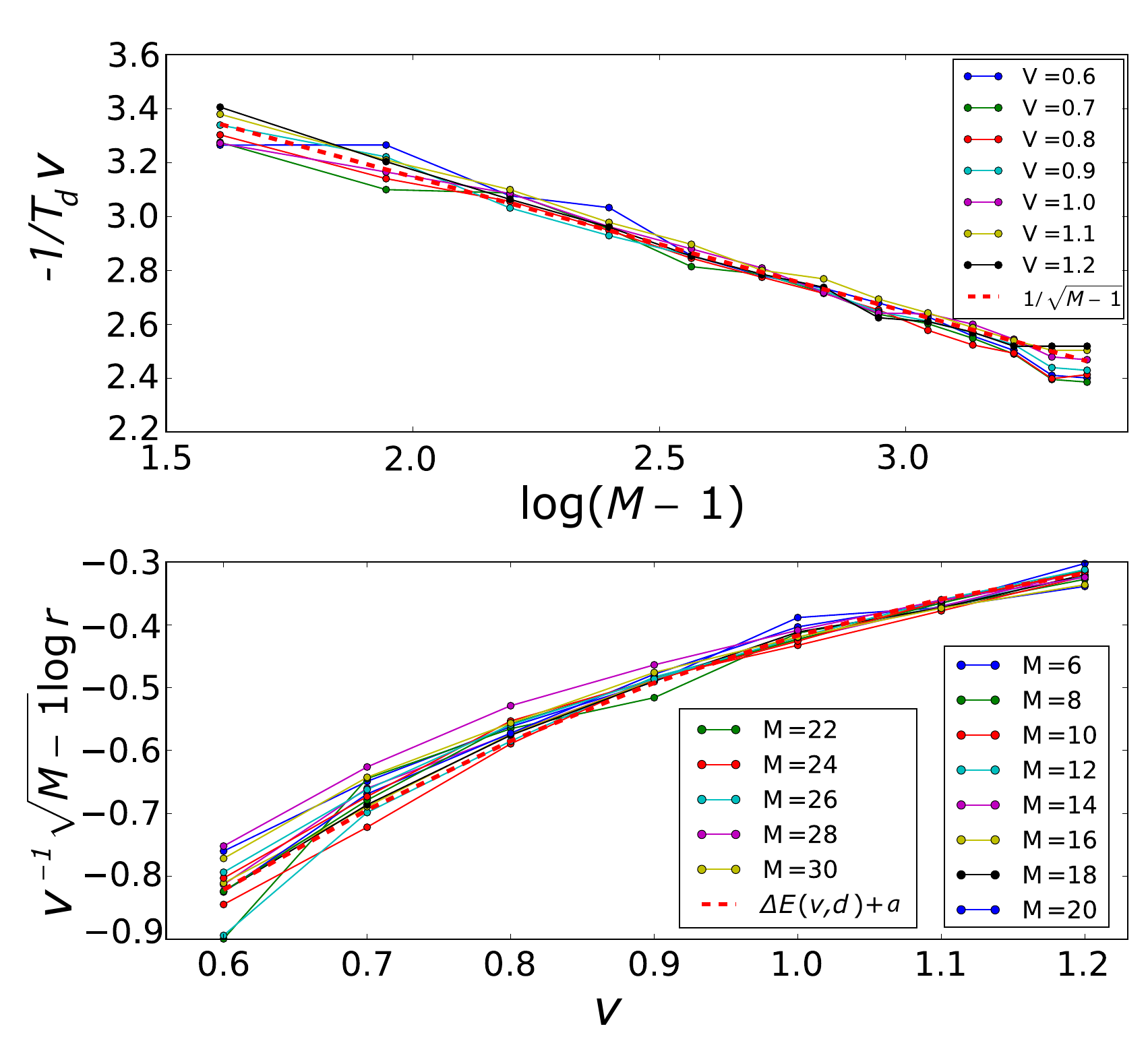}
	\caption{\textbf{Top}: Plotting $-1/T_d v = \log\{ v^{-1} \log r /[\Delta E(v,d) + a] \}$ against $\log (M-1)$. Different solid lines corresponds to data with different $v$, and the dashed line corresponds to the $(M-1)^{-1/2}$ curve. \textbf{Bottom}: Plotting $v^{-1} \log r \sqrt{M-1} \propto \Delta E(v,d)$ against $v$. Different solid lines corresponds to data with different $M$, and dashed line corresponds to the $\Delta E(v,d)+a$ curve. \label{fig:barrier_collapse}}. 
\end{figure}

Therefore, our model Eqn.\eqref{eqn:pretrieval} has in total three parameters to determine $\gamma$, $k$, and $a$. In Fig.\ref{fig:barrier_collapse} we determine the parameters by collapsing data (see details of the collapse in below and in caption), and find that the best fit is found provided $\gamma = 240.30, k = 5255.0k_B^{-1}, a = -0.35445$. Henceforth we fix these three parameters to these values.

In Fig.\ref{fig:barrier_collapse} bottom, we offset the effect of $M$ by multiplying $v^{-1}\log r$ by $\sqrt{M-1}$, and we see that curves corresponding to different $M$ collapse to each other, confirming the $\sqrt{M-1}$ dependence in $T_d$. The collapsed line we are left with is just the $v$-dependence of $\Delta E(v,d)$, up to overall constant.

In Fig.\ref{fig:barrier_collapse} top, we offset the effect of $v$ in $T_d$ by multiplying $v^{-1}$ to $ \log r /[\Delta E(v,d) + a]$. We see that different curves corresponding to different $v$'s collapse to each other, confirming the $v^{-1}$ dependence in $T_d$. The curve we are left with is the $M$ dependence in $T_d$, which we see fits nicely with the predicted $\sqrt{M-1}$.  

In main text Fig.\ref{fig:capacity}(b) we run our simulation with the following parameters held fixed: $N=4000, \; p=10, \; \epsilon=0.35, \; \; \tau=1, \; J=100, \; d=10, \; w=30$. Along the same curve, we vary $M$ from $6$ to $30$, and the series of curves corresponds to different $v$ from $0.6$ to $1.2$. 

In  main text Fig.\ref{fig:capacity}(c) we hold the following parameters fixed: $p=10,\;  \epsilon=0.35, \; \tau=1, \; J=100, \; d=10, \; w=30, \; v=0.8$. Along the same curve, we vary $M/\frac{N}{(\log N)^2}$ from $0.1$ to $0.6$, and the series of curves corresponds to different $N$ from $1000$ to $8000$. 

In both  main text Fig.\ref{fig:capacity}(b)(c) the theoretical model we used is Eqn.\eqref{eqn:pretrieval} with the same parameters given above.

In main text Fig.\ref{fig:capacity}(d) we re-plot the theory and data from main text Fig.\ref{fig:capacity}(b) in the following way: for the theoretical curve, we find the location where $P_{retrieval} = 0.5$, and call the corresponding $M$ value theoretical capacity; for the simulation curve, we extrapolate to where $P_{retrieval} = 0.5$, and call the corresponding $M$ value, the simulation capacity.

For all simulation curves above, we drag the droplet from one end of the continuous attractor to the other end of the attractor, and run the simulation for 300 times. We then measure the fraction of successful events (defined as the droplet survived in the cup throughout the entire trajectory of moving) and failed events (defined as the droplet escape from the cup at some point before reaching the other end of the continuous attractor). We then define the simulation $P_{retreival}$ as the fraction of successful events.

\end{document}